\documentclass[amsmath,amssymb,amsbsy,prl,twocolumn,superscriptaddress,showpacs]{revtex4}
\usepackage{amsfonts}
\usepackage{amsmath}
\usepackage{amssymb}
\usepackage{bm}
\usepackage[usenames, dvipsnames]{color}
\usepackage{dcolumn}
\usepackage{epsfig}
\usepackage{eucal}
\usepackage{graphicx}
\usepackage[breaklinks,colorlinks = true,linkcolor = red,urlcolor=cyan,citecolor=red]{hyperref}
\usepackage{mathrsfs}
\usepackage{subfigure}
\usepackage{textcomp}
\usepackage{times}
\usepackage{xy}
\usepackage{color}

\usepackage[toc,page]{appendix}

\begin{document}

\title{Coupling between improper ferroelectricity and ferrimagnetism  in hexagonal ferrites}
\author{Hena Das}
\email{das.h.aa@m.titech.ac.jp}
\affiliation{Laboratory for Materials and Structures, Tokyo Institute of Technology, 4259 Nagatesuta, Midori-ku, Yokohama, Kanagawa 226-8503, Japan}
\affiliation{Tokyo Tech World Research Hub Initiative (WRHI), Institute of Innovative Research, Tokyo Institute of Technology, 4259 Nagatsuta, Midori-ku, Yokohama, Kanagawa 226-8503, Japan}
\date{\today}

\begin{abstract}
Antisymmetric Dzyaloshinskii-Moriya (DM) interactions generating from the spin-orbit coupling induce various fascinating properties, like magnetoelectric (ME) effect, weak ferromagnetism and non-trivial topological spin textures like skyrmions, in real materials. Compared to their symmetric isotropic exchange counterpart, these interactions are generally of a weaker order of strength, creating modest twisting in the spin structure which results in weak ferromagntism or weak linear ME effect. Our proposed two-sublattice model, in contrast, predicts a hitherto unobserved, charge ordered non-collinear ferrimagnetic behavior with a considerably high magnetization $\textbf{M}$ coexisting with a ferroelectric (FE) order with an electric polarization $\textbf{P}$ and a strong cross coupling between them which is primarily driven by the inter-sublattice DM interactions. The key to realize these effects is the coupling between these microscopic interactions and the FE primary order parameter. We predict microscopic mechanisms to achieve electric field $\textbf{E}$ induced spin-reorientation transitions and 180$^{\circ}$ switching of the direction of $\textbf{M}$. This model was realized in the hexagonal phase of LuFeO$_3$ doped with electrons. This system shows $P \sim$ 15 $\mu$C/cm$^2$, $M \sim$ 1.3 $\mu_B$/Fe and magnetic transition near room temperature ($\sim$ 290 K). Our theoretical results are expected to stimulate further quest for energy-efficient routes to control magnetism for spintronics applications.

\end{abstract}

\maketitle
\noindent
Multisublattice magnets show various fascinating  properties, like spin-reorientation (SR) transitions~\cite{RFO-MAG, ref2, yamaguchi, tokura, Rao, Tokunaga, Zhao, DAS}, non-trivial ferrimagnetism and topological order ~\citep{FIM1,FIM2,FIM3,FIM4,FIM5,FIM6,FIM7,FIM8,FIM9}, which owe their origin to various microscopic interactions. Moreover, the quest for efficient routes of manipulation of these microscopic interactions by the application of external stimuli in order to control the properties of these systems, is at the forefront of various research activities~\cite{FIM9,FIM10,FIM11,FIM12,FIM13,FIM14}. LuFe$_2$O$_4$ is one such system, exhibiting Fe$^{2+}$/Fe$^{3+}$ charge ordered (CO) pattern in the Fe double layer, resulting in ferrimagnetic order with a robust magnetization $\textbf{M} \sim$ 0.8 - 1.4 $\mu_B$/Fe~\cite{LFO124-MAG1} below $\sim$ 240 K~\citep{LFO124-1,LFO124-2} and to the genesis of various interesting functionalities~\citep{LFO124-F1,LFO124-F2,LFO124-F3,LFO124-F4}, though its ferroelectric (FE) order remains ambiguous~\cite{LFO124-1,LFO124-2,LFO124-3,LFO-AFE1,LFO-AFE2,LFO-AFE3,LFO-AFE4,LFO-JM1}. 
The constructed (LuFeO$_3$)$_m$/(LuFe$_2$O$_4$)$_1$ superlattices manifest various interesting properties~\citep{LFO-JM1,LFO-JM,LFO-JM2}, like room temperature multiferroic and magnetoelectric (ME) behavior and dimensionality controlled topological order. Hexagonal LuFeO$_3$ shows improper FE behavior with an electric polarization ($\textbf{P}$) below $\sim$ 1040 K and canted antiferromagnetic (AFM) behavior below $\sim$ 147 K with a net magnetization of $\sim$ 0.03 $\mu_B$/Fe~\citep{LFO113-1,LFO113-HD,LFO113-2,LFO113-3, LFO113-4}. The ferrimagnetic magnetization ( $\textbf{M}$ ) in LuFe$_2$O$_4$ arises due the CO pattern ($C_\textbf{q}$) characterized by the wave vector $\textbf{q}=(\frac{1}{3},\frac{1}{3},\frac{\eta}{3})$, while in LuFeO$_3$ $\textbf{P}$ gets induced by an zone boundary structural distortion ( $\textbf{Q}_{K_3}$ ) following $K_3$ symmetry at $\textbf{k}=(\frac{1}{3},\frac{1}{3},0)$ of the paraelectric (PE) $P6_3/mmc$ structure~\citep{LFO-FE1,LFO-FE2,LFO-FE3}. The engineered superlattices not only exhibit enhancement in the magnetic transition temperature and magnetization compared to the parent systems, but are also reported to show strong ME switching phenomena~\cite{LFO-JM1}. However, the cross-coupling between ferrimagnetism and ferroelectricity is not well understood yet.


Here, employing first-principles density functional theory (DFT) calculations and finite temperature Monte Carlo (MC) simulations, we propose an alternate model where such C$_\textbf{q}$-type CO pattern forms in the LuFeO$_3$ structure itself, under electron doping. 
We observed that the doped system retains its improper FE nature, with an electric polarization $\textbf{P}$ at par with the parent system. The C$_\textbf{q}$-type pattern creates a two-sublattice model where the Fe$^{2+}$ triangular lattice is embedded within the Fe$^{3+}$ hexagonal lattice, resulting in the formation of unique non-collinear ferrimagnetic phases with the spins in the two magnetic sublattices aligned in mutually perpendicular ($\perp$) directions, giving rise to a high magnetization $\textbf{M}$. 
As in ferrite superlattices, the order of magnitude of $\textbf{M}$ and the corresponding magnetic transition temperature, both show significant enhancement in comparison to the parent material. These phases originate primarily due to the interplay between inter-sublattice Dzyaloshinskii-Moriya (DM) interactions and $\textbf{Q}_{K_3}$, predicting microscopic mechanisms to achieve electric field induced SR transitions and 180$^{\circ}$ ME switching, where $\textbf{P}$ and $\textbf{M}$ simultaneously switch their individual orientations.


\begin{figure}[h]
\begin{center}
\includegraphics[scale=0.3]{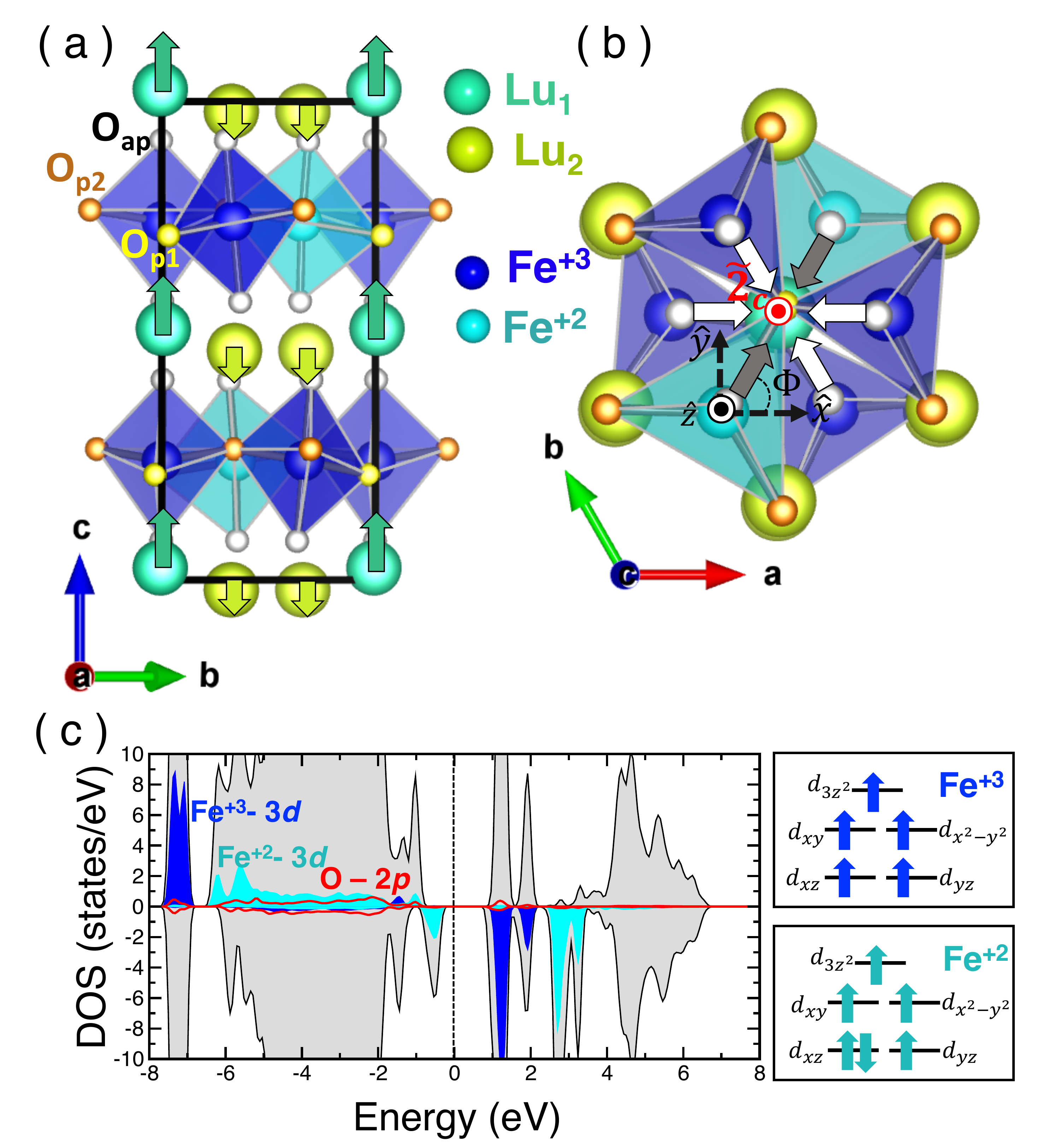} 
\end{center}
\caption{$Cmc2_1$ crystal structure doped LuFeO$_3$ exhibiting $\textbf{Q}_{K_3}$ distortions consist of (a) one-up/two-down buckling (denoted with arrows) of the Lu layer and (b) tilting of the FeO$_5$ trigonal bipyramids (represented by the arrows) towards the two-fold screw axis $\tilde{2}_c$. The phase factor $\Phi$ is defined with respect to the Cartesian coordinate. (c) Right panel: calculated density of states (DOS) of the $e^{-}$-doped LuFeO$_3$. Left panel: electron occupancy of the Fe$^{3+}$ and Fe$^{2+}$ ions.The trigonal bipyramid oxygen environment splits the 3$d$ level of the Fe ions as, $e^{\prime}$ ($d_{xz}$ and $d_{yz}$), $e^{\prime\prime}$ ($d_{xy}$ and $d_{x^2-y^2}$) and $a^{\prime}_1$ ($d_{3z^2}$), in the order of increasing energy.}
\label{QK3}
\end{figure}

In LuFeO$_3$, the FE order is improper as the $\textbf{P}$ is the secondary effect of a primary distortion, $\textbf{Q}_{K_3}$.
The $\textbf{Q}_{K_3}$ distortion, which breaks the inversion symmetry of $P6_3/mmc$, assumes the form of tilted FeO$_5$ bipyramids towards (away) the $\tilde{2}_c$ axis and results in one-up/two-down (one-down/two-up) buckling of the Lu ions~\cite{LFO113-HD}. $\textbf{Q}_{K_3}$ is defined by its magnitude $Q_{K_3}$ and a phase factor $\Phi = \frac{n\pi}{3}$ representing the direction of the tilt of FeO$_5$ bipyramids. The non-linear coupling between $\textbf{Q}_{K_3}$ and $\textbf{Q}_{\Gamma^{-}_2}$ induces electric polarization $\textbf{P}$ directed along the crystallographic $\textbf{c}$ ($\hat{z}$) axis, following the Landau free energy~\cite{LFO113-HD,LFO-FE3}, 
\begin{equation}
\mathcal{F} \sim PQ^{3}_{K_3}cos(3\Phi)
\end{equation}
where $n$ takes up six discrete values ($n=1,2,3,4,5,6$) due to $Z_6$ symmetry, where $\textbf{P}$ of the odd and even number structures are oriented in mutually opposite directions~\cite{LFO-NS}. This leads to the formation of topologically protected FE and ME vortex domain structures in rare-earth manganites and ferrites~\citep{TPD-1,TPD-2,TPD-3,TPD-YU,TPD-4,TPD-5,TPD-6, LFO-JM2}. 
Notably $P\bar{3}c1$ also transforms as $K_3$ symmetry (see Sec. II in Supplementary Materials~\cite{SM}).
GGA+$U$ results (Sec. I in the Supplementary Materials~\citep{SM}) considering $U =$ 4.5 eV and $J_H = $ 0.95 eV at the Fe sites~\citep{LFO-JM1,LFO-JM,LFO113-HD}, show that LuFeO$_3$ crystallizes in the FE $P6_3cm$ phase with 24 meV/f.u. lower in energy compared to $P\bar{3}c1$ structure. It exhibits $P \sim$ 14 $\mu$C/cm$^2$ which agrees with the previous reports~\cite{LFO113-HD}. 

\begin{figure}[h]
\begin{center}
\includegraphics[scale=0.32]{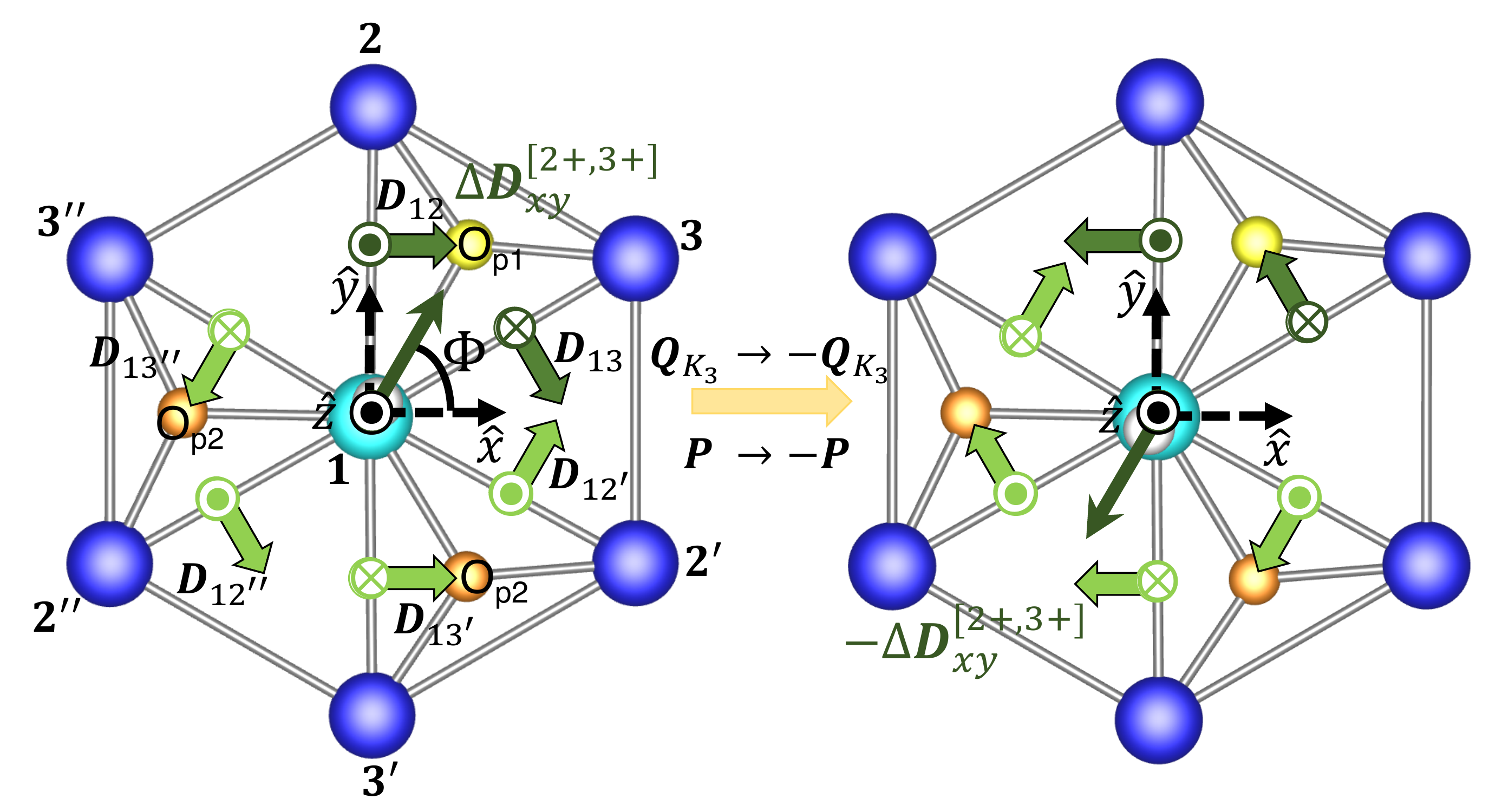} 
\end{center}
\caption{NN DM vectors acting on the Fe 1 for a single Fe layer. The longitudinal component of the DM vectors orientated along the positive and negative $\hat{z}$ directions are shown with dot and cross marks, respectively. The $\textbf{Q}_{K_3}$ distortions induce two non-equivalent transverse components of DM vectors mediated through planar oxygens O$_{p1}$ ($D_{12}= D_{13}=D$) and O$_{p2}$ ($D_{12^{\prime}} = D_{13^{\prime}} = D_{12^{\prime\prime}} = D_{13^{\prime\prime}} = D^{\prime}$), respectively. $\Delta \textbf{D}_{xy}^{[2+,3+]}$ denotes effective DM interaction acting on Fe 1 as derived in Eq.~\ref{D}. The transverse components switches their direction with the $\textbf{Q}_{K_3}\rightarrow-\textbf{Q}_{K_3}$ ($\textbf{P} \rightarrow -\textbf{P}$) switching process.}
\label{model}
\end{figure}

\begin{figure*}
\begin{center}
\includegraphics[scale=0.37]{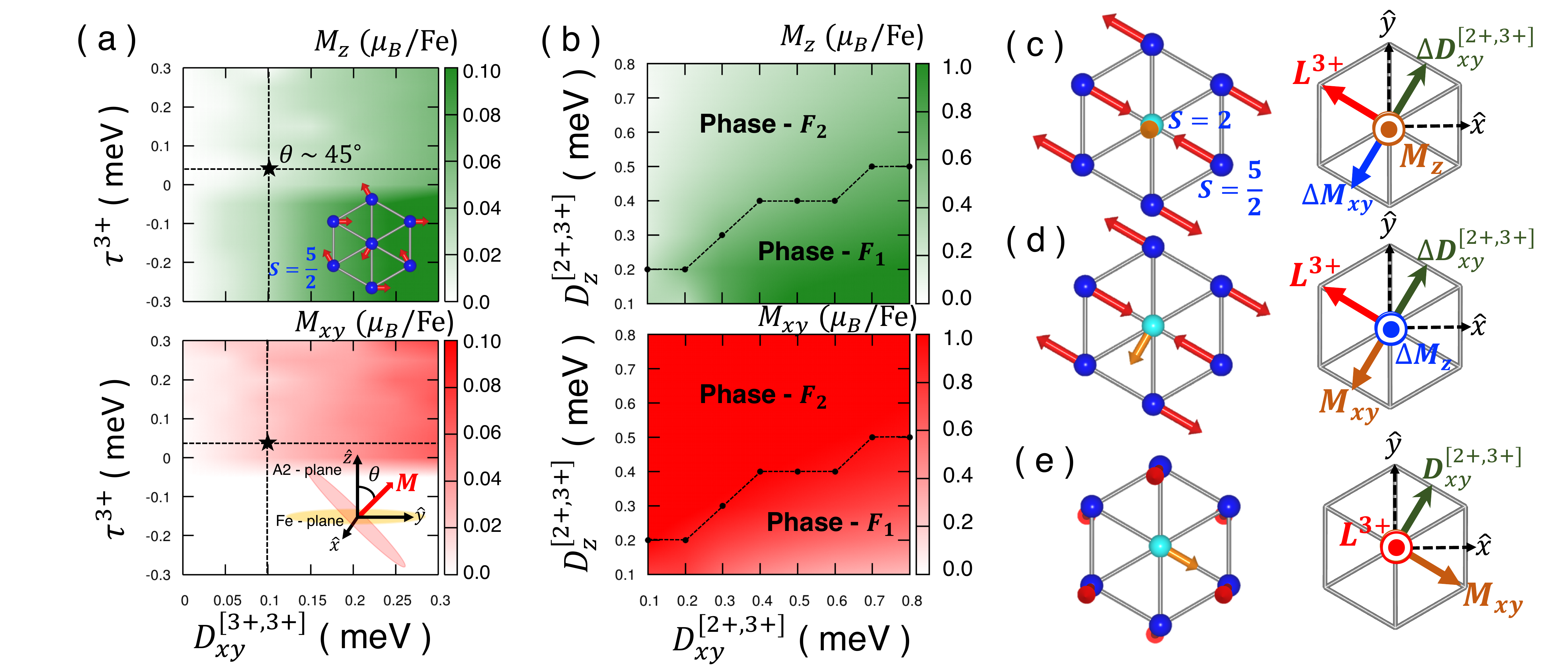} 
\end{center}
\caption{( a ) Calculated out-of-plane $M_z$ (upper panel) and in-plane $M_{xy}$ (lower panel) component of magnetization ($\textbf{M}$) as functions of $D_{xy}^{[3+,3+]}$ and SIA parameter $\tau^{3+}$, employing $H_1$ model. Inset; A2 spin order and the tilt angle ($\theta$) of $\textbf{M}$ with respect to $\hat{z}$ axis. Star denotes DFT estimated values of $D_{xy}^{[3+,3+]}$ and $\tau^{3+}$. The results correspond to $D_z^{[3+,3+]} ~\sim$ 0.1 meV as estimated from DFT~\cite{LFO113-HD}. ( b ) Calculated out-of-plane $M_z$ (upper panel) and in-plane $M_{xy}$ (lower panel) magnetization as functions of $\textbf{D}^{[2+,3+]}$ parameters, using $H_2$ model. ( d ) - ( f ), identified non-collinear ferrimagnetic orders $F_1$, $F_2$ and $F_3$, respectively and their schematic representations. Red and orange arrows represent Fe$^{3+}$ and Fe$^{2+}$ spins, respectively.}
\label{MC1}
\end{figure*}

We indroduced excess carriers in the system by changing the number of valence electrons and adding a homogeneous background charge to keep the system neutral, like in previous studies~\cite{e-FE1,e-FE2}. Optimizing both doped $P6_3cm$ and $P\bar{3}c1$ structures, we observed that, LuFeO$_3$ at the $x = \frac{1}{3}$ electrons ($e^{-}$) per Fe doping level clearly shows the formation two distinct type of Fe ions (see Fig.~\ref{QK3}). Where $\frac{2}{3}$ of them show completely filled majority spin channels leaving the minority channels almost empty and hence a 3+ nominal oxidation state (3$d^5$ configuration), similar to LuFeO$_3$~\cite{LFO113-HD}. Whereas $\frac{1}{3}$ of them show partial occupancy in the minority spin channels (with 3$d^6$ electronic configuration resulting in 2+ oxidation state) in addition to completely filled majority spin channels. A $C_\textbf{q}$-type CO $Cmc2_1$ structure with each Fe layer consisting of Fe$^{2+}$ triangular lattice embedded within Fe$^{3+}$ hexagonal lattice (Fig.~\ref{QK3}(a) and (b)) is formed, as was observed in LuFe$_2$O$_4$. The doped system exhibits $\textbf{Q}_{K_3}$ distortion characterized by an electric polarization $P \sim$ 15 $\mu$C/cm$^2$ and insulating behavior with $\sim$ 0.8 eV band gap, similar to parent LuFeO$_3$. The mixed valent state was 180 meV/f.u. higher in energy (see Sec. II in the Supplementary Materials~\cite{SM} for details). 
%

To determine the resulting magnetic order of the CO two-sublattice system, we conducted MC simulations considering spin Hamiltonian with the general form,
\begin{eqnarray}
H = \sum_{i\neq j} J_{ij}\textbf{S}_i\cdot \textbf{S}_j+\sum_{i\neq j}\textbf{D}_{ij}\cdot \textbf{S}_i\times \textbf{S}_j+\sum_i \textbf{S}_i\cdot \hat{\tau}_i\cdot \textbf{S}_i
\end{eqnarray}
here $J_{ij}$ and $\textbf{D}_{ij}$ represent the Fe-Fe symmetric exchange (SE) and antisymmetric (DM) interactions, respectively. $\hat{\tau}_i$ denotes single ion anisotropy (SIA) tensor of the Fe ions. We constructed two models, $H_1$ and $H_2$, corresponding to the undoped (one-sublattice ) and doped (two-sublattice) systems, respectively (detail description of these models and subsequent MC simulations are given in Sec. III of the Supplementary Materials~\cite{SM}). $H_1$ and $H_2$ consist of, ( i ) nearest-neighbor (NN), second-nearest-neighbor (2NN) in-plane and effective inter-layer Fe-Fe SE interactions. ( ii ) NN Fe-Fe DM interactions as depicted in Fig.~\ref{model}. The transverse component is induced by the distortion $\textbf{Q}_{K_3}$, while the longitudinal component is already existing in the PE phase. The DM vectors in the consecutive Fe layers are anti-parallel to each other due to $\tilde{2}_c$ symmetry. (iii) The SIA $\hat{\tau}$ has $\tau_{xx} \neq \tau_{yy} \neq \tau_{zz}$ and non-zero off-diagonal components $\tau_{xz}=\tau_{zx}$.

Fig.~\ref{MC1}(a) shows the results of MC simulations employing model $H_1$. 
The estimated values of DM interactions and SIA parameters (Sec.III-B in the supplementary Materials~\citep{SM}) show that, \textit{i.e.} $D \sim D^{\prime}=D^{[3+,3+]}$ (Fig.~\ref{model}) and $\tau_{xx} \sim \tau_{yy} = \tau^{3+}$. We incorporated these postulates in MC simulations. Both NN and 2NN interactions are AFM in nature with estimated strengths of $J_{NN}^{[3+,3+]} \sim$ 6.3 meV and $J_{2NN}^{[3+,3+]} \sim$ 0.3 meV, respectively, agreeing well with the previous reports~\citep{LFO113-HD}.  
Geometric frustration created by the six NN AFM Fe-Fe SE interactions in the triangular lattice induces the formation of multiple, energetically degenerate, 120$^{\circ}$ non-collinear orders~\citep{LFO113-HD}. 
The effective AFM $\bigtriangleup J_c \sim$ 0.4 meV interaction breaks the geometric frustration and stabilizes the A2-type magnetic order ($P6_3c^{\prime}m^{\prime}$) (Fig.~\ref{MC1}(a)). 
Notably, Fe$^{3+}$ ions result in uniaxial ($\hat{z}$) magnetic anisotropy, contrasting the in-plane ($xy$) magnetic anisotropy of Mn$^{3+}$ ions in its manganite counterpart~\cite{LFO113-HD}, which tilts the A2 spin ordered plane (Fig.~\ref{MC1}(a)). $D_{xy}^{[3+,3+]}$ gives rise to a modest canted magnetization. DFT estimated values of magnetic parameters stabilize $\theta \sim $ 45$^{\circ}$ tilted A2 phase below $\sim$ 148 K exhibiting a canted magnetization of $\sim$ 0.03 $\mu_B$/Fe (Fig.S5 and S6 in the Supplementary Materials~\cite{SM}). These observations are at par with experimental reports~\cite{LFO113-1,LFO113-HD,LFO113-2,LFO113-3, LFO113-4}, establishing the power of this approach.


Fig.~\ref{MC1}(b) shows results of MC simulations employing model $H_2$. 
The key modifications of $H_2$ compared to $H_1$ are, (1) the strength of the NN Fe$^{3+}$-Fe$^{3+}$ interaction $J_{NN}^{[3+,3+]}$ is increased from 6.3 $\rightarrow$ 8.5 meV (primarily due to the enhancement of the $\angle$Fe$^{3+}$-O-Fe$^{3+}$ of the mediating path (Fig. S8 in Supplementary Materials~\cite{SM}). (2) The NN AFM Fe$^{2+}$-Fe$^{3+}$ interaction $J_{NN}^{[2+,3+]}$ $\sim$ 1.1 meV is introduced. 
(3) A weak FM 2NN Fe$^{2+}$-Fe$^{2+}$ interaction $J_{2NN}^{[2+,2+]} \sim $ -0.1 meV is incorporated.
(4) Fe$^{2+}$ ions exhibit strong spin-orbit (LS) coupling with orbital moment $\mu_{o}^{2+} \sim$ 0.2 $\mu_B$, order of magnitude stronger than Fe$^{3+}$ ion ($\frac{ \mu_{o}^{3+}}{\mu_{o}^{2+}} \sim$ 0.1). As in LuFe$_2$O$_4$~\citep{LFO124-F1,LFO124-F2,LFO124-F3,LFO124-F4}, Fe$^{2+}$ uniaxial magnetic anisotropy having $\tau ^{2+} \sim $ 0.2 meV was included ($\frac{\tau^{3+}}{\tau^{2+}} \sim 0.1$). (5) The inter-sublattice DM interactions ($\textbf{D}^{[2+,3+]}$), are likely to be stronger than their intra-sublattice counterparts ($\textbf{D}^{[3+,3+]}$), playing a significant role in the determination of the magnetic order. 

\begin{figure}
\begin{center}
\includegraphics[scale=0.22]{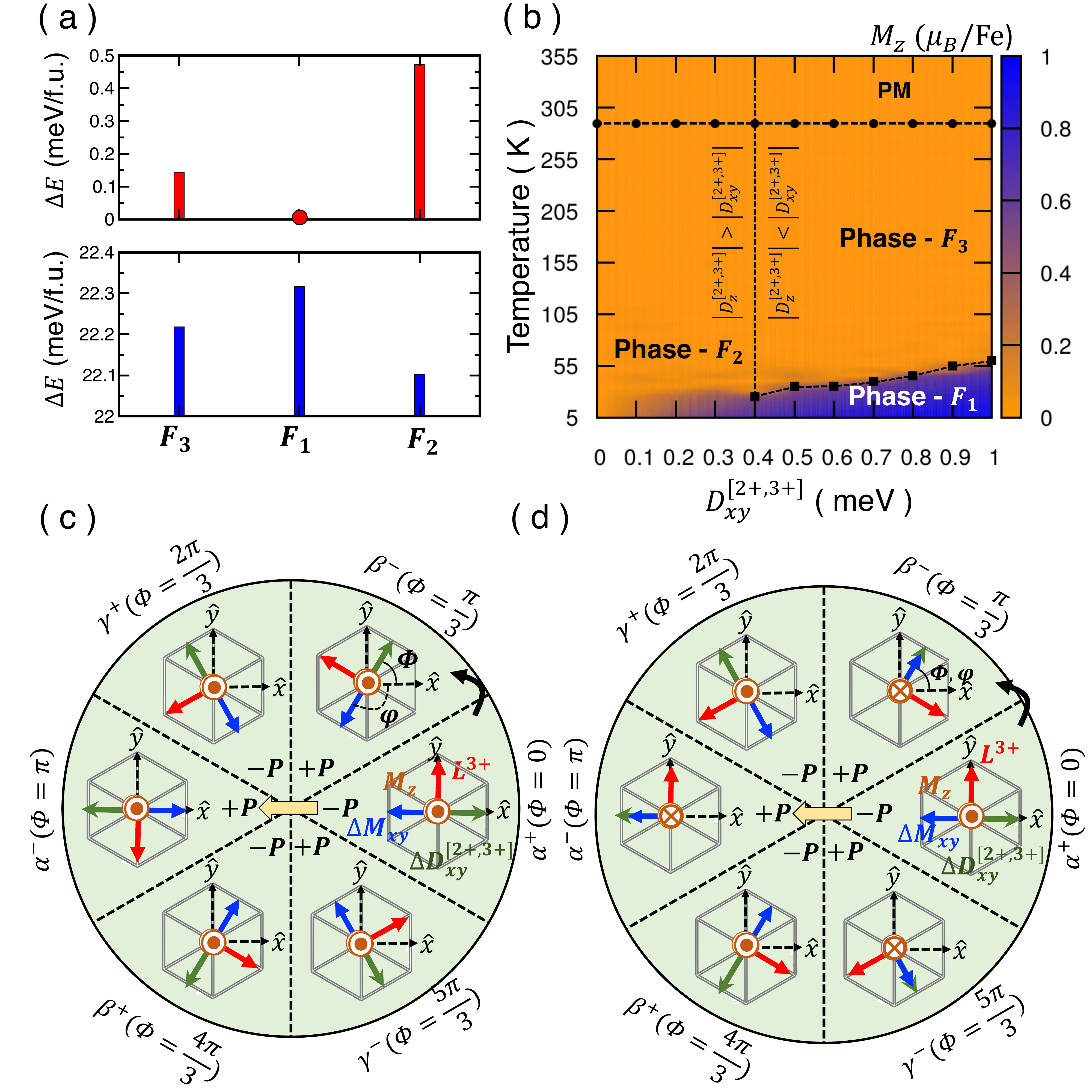} 
\end{center}
\caption{( a ) GGA+$U$ calculated total energy of the ferrimagnetic phases in polar (upper panel) and non-polar (lower panel) structures with respected to the energy of polar $F_1$ phase. ( b ) constructed temperature vs $D_{xy}^{[2+,3+]}$ phase diagram using MC results. PM represents paramagnetic phase. The PM to ferrimagnetic order and SR transition temperatures are marked with solid black spheres and squares, respectively. ( c ) Probable ME domain structures.}
\label{MC2}
\end{figure}

Our results (Fig.~\ref{MC1}(b)) show the formation of non-collinear ferrimagnetic orders, where the FM ordered Fe$^{2+}$ spins are $\bot$ oriented with respect to the AFM ordered Fe$^{3+}$ sublattice ($\textbf{L}^{3+}$ denotes AFM order parameter), resulting in a net magnetization $\textbf{M}$. Without $\textbf{D}^{[2+,3+]}$, the Fe$^{2+}$ spins do not show any cooperative order. Three magnetic phases were identified, namely $F_1$, $F_2$ and $F_3$ (Fig.~\ref{MC1}(c)-(e)). $F_1$ aligns the major components of the Fe$^{2+}$ and Fe$^{3+}$ spins along the $\hat{z}$ axis and in the $xy$ plane, respectively, giving rise to longitudinal and transverse components of $\textbf{M}$ as,
\begin{equation}
\textbf{M}_z \propto \Delta \textbf{D}_{xy}^{[2+,3+]} \times \textbf{L}^{3+}\\
\label{M1}
\end{equation}
\begin{equation}
\Delta\textbf{M}_{xy} \propto \Delta \textbf{D}_{z}^{[2+,3+]} \times \textbf{L}^{3+}
\label{M2}
\end{equation}
$F_2$ corresponds to co-planar mutually $\perp$ magnetic order with major $\textbf{M}_{xy}$ and minor $\Delta \textbf{M}_{z}$ (induced by $\Delta\textbf{D}_{xy}^{[2+,3+]}$) components. In $F_3$, the major Fe$^{2+}$ and Fe$^{3+}$ spins are oriented in ${xy}$ plane and along the $\hat{z}$ axis, respectively. The stability of these phases is controlled by the complex interplay between the effective Fe$^{2+}$-Fe$^{3+}$ DM interactions, 
\begin{equation}
\Delta\textbf{D}^{[2+,3+]} \approx (\bar{D}_{xy}^{[2+,3+]}cos\Phi,\bar{D}_{xy}^{[2+,3+]}sin\Phi,\bar{D}_{z}^{[2+,3+]})
\label{D}
\end{equation}
and the SIA ($\tau^{2+}$ and $\tau^{3+}$). Here, $\bar{D}_{xy}^{[2+,3+]} = 2D_{xy}^{[2+,3+]}$ and $\bar{D}_{z}^{[2+,3+]}=6D_{z}^{[2+,3+]}$. Both the direction and magnitude of $\bar{D}_{xy}^{[2+,3+]}$ are synchronized with $\textbf{Q}_{K_3}$ (Fig.~\ref{model} and Sec. IV of the Supplementary Materials~\cite{SM}).

Next, we conducted GGA+$U$ total energy calculations considering the polar and non-polar structures of the ferrimagnetic orders. 
The $F_1$ and $F_2$ ground state magnetic orders appear in the polar and the non-polar phases, respectively (Fig.~\ref{MC2}(a)). These results are in harmony with the MC solutions obtained in the $D_{xy}^{[2+,3+]} > D_{z}^{[2+,3+]}$ magnetic parameter space (Fig.~\ref{MC2}(b)). This indicates $F_1 \rightarrow F_2$ SR transition with the modulation on the $\Delta\textbf{ D}_{xy}^{[2+,3+]}$ parameter synchronized with $\textbf{Q}_{K_3}$. 
Temperature also drives $F_1 \rightarrow F_3$ SR transition (Fig.~\ref{MC2}(b)). At low temperature ( $\sim$ 5 K) $F_1$ cooperative order forms as $\frac{\tau^{2+}}{\tau^{3+}}>\frac{2(S^{3+})^2}{(S^{2+})^2}$. However, at high temperature $F_3$ order is formed due to entropy.
%
%
The paramagnetic (PM) to ferrimagnetic phase transition temperature ($T_c$) is primarily determined by the relative $\frac{J_{NN}^{[2+,3+]}}{J_{NN}^{[3+,3+]}}$ strength, which also controls the magnitude of magnetization (Fig.S11 in the Supplementary Materials~\citep{SM}). 
In the $e^-$ doped system, $T_c \sim $ 290 K and  $M \sim$ 1.3 $\mu_B$/Fe(see Fig.S12 in the Supplementary Materials~\citep{SM}), both are significantly higher than their LuFeO$_3$ counterparts~\citep{LFO113-1,LFO113-HD,LFO113-2,LFO113-3, LFO113-4}, indicating an effective route to enhance the magnetic properties of LuFeO$_3$.

The manipulation of $\textbf{P}$ in these systems is complex in nature and not fully understood yet. However, the non-llinear coupling between $\textbf{P}$ and $\textbf{Q}_{K_3}$ is a well established fact~\citep{LFO113-HD,LFO-FE3}. 
The $\textbf{P} \rightarrow -\textbf{P}$ process leads to either $\textbf{Q}_{K_3} \rightarrow -\textbf{Q}_{K_3}$ or rotation of $\textbf{Q}_{K_3}$ by an angle $\Delta \Phi = \frac{\pi}{3}$ (Fig.~\ref{MC2}(c) and (d)). The tri-linear coupling between $\Delta \textbf{D}^{[2+,3+]}$, $\textbf{M}$ and $\textbf{L}^{3+}$ (Eq.~\ref{M1} and Eq.~\ref{M2}) and the coupling between $\Delta \textbf{D}^{[2+,3+]}$ and $\textbf{Q}_{K_3}$ (Eq.~\ref{D} and Fig.~\ref{model}) show that,
 $\textbf{Q}_{K_3} \rightarrow -\textbf{Q}_{K_3}$ ($\alpha^+ \rightarrow \alpha^-$) either reverses the direction of $\textbf{L}^{3+}$ (reversing the direction of $\textbf{M}_{xy}$) or  that of $\textbf{M}_z$. 
Moreover, formation of the 'clover-leaf' vortex domain pattern~\citep{TPD-1,TPD-2,TPD-3,TPD-YU,TPD-4,TPD-5,TPD-6}, suggests two probable ME domain structures. The first is characterized by the rotation of $\textbf{Q}_{K_3}$ to its NN domain ($\Delta\Phi=\pm\frac{\pi}{3}$) leading both $\textbf{L}^{3+}$ and $\Delta\textbf{M}_{xy}$ to rotate in phase by $\frac{\pi}{3}$. However, $\textbf{M}_z$ does not change its direction (Fig.~\ref{MC2}(c)). In the second structure, the rotation of $\textbf{Q}_{K_3}$ to its NN domain ($\Delta\Phi=\pm\frac{\pi}{3}$) induces both $\textbf{L}^{3+}$ and $\Delta\textbf{M}_{xy}$ to rotate out of phase by $\frac{2\pi}{3}$ leading to the 180$^{\circ}$ switching of $\textbf{M}_z$ (Fig.~\ref{MC2}(d). In the present system, $\textbf{M}_z \rightarrow -\textbf{M}_z$ process is expected to be more feasible, as the intermediate $F_2$ state exhibits lowest transition barrier height (Fig.~\ref{MC2}(a) and (b)) and is associated with lower magnetostatic energy. 
This phenomenon opens up effective routes to achieve 180$^{\circ}$ ME switching via electric polling method~\citep{TPD-2}.

In (LuFeO$_3$)$_m$/(LuFe$_2$O$_4$) superlattices, previous DFT calculations and EELS measurements showed the formation of the hole-doped LuFe$_2$O$_4$ ferrimagnetic tail-to-tail domain wall~\citep{LFO-JM1,LFO-JM2}, transferring $e^-$ to the LuFeO$_3$ layer forming head-to-head wall. Also, a coupling between $\textbf{Q}_{K_3}$ and $T_c$ based on LuFe$_2$O$_4$ FE model was proposed, where an increase in the former led to a subsequent increase in the latter~\citep{LFO-JM1}. Our present study, on the other hand, proposes $e^-$ doped LuFeO$_3$ FE model, where localization of the doped $e^-$ in the form of C$_\textbf{q}$-type order and the formation of $F_1$ ferrimagnetic order can lead to a significant enhancement in $\textbf{M}$ ($\sim$ 2.4 - 4.0 $\mu_B$ per LuFe$_2$O$_4$ f.u.)~\citep{LFO-JM1} and a likely increase in $T_c$. Additionally, it is expected to contribute to a complex ME switching process which is worth exploration.

In summery, we predict non-collinear ferrimagnetism, SR transitions and 180$^{\circ}$ ME switching phenomena governed by the coupling between charge ordering, DM interactions and improper ferroelectricity. These predictions are based on MC simulations on a two-sublattice model, constructed by doping the improper FE hexagonal phase of LuFeO$_3$ with electrons. We elucidate the prospective microscopic mechanisms to control the stabilization of ferrimagnetic orders by applying electric field $\textbf{E}$ induced SR transitions and 180$^{\circ}$ switching of the direction of $\textbf{M}$. Our proposed model will expectedly motivate the designing of non-collinear ferrimagnetic and ME materials with prospective applications in spintronics technology.

H.D. acknowledges the fruitful discussions with Saurabh Ghosh, M.J.Swamynadhan, Andrew O'Hara and Sokrates T. Pantelides. Research at the Tokyo Institute of Technology is supported by the Grants-in-Aid for Scientific Research No. 19K05246 from the Japan Society for the Promotion of Science (JSPS). H.D. also acknowledges computational support from TSUBAME supercomputing facility.

\newcommand{\beginsupplement}{%
        \setcounter{table}{0}
        \renewcommand{\thetable}{S\arabic{table}}%
        \setcounter{figure}{0}
        \renewcommand{\thefigure}{S\arabic{figure}}%
        \setcounter{equation}{0}
        \renewcommand{\theequation}{S\arabic{equation}}%
     }

\beginsupplement

\section{Supplementary Materials}

\section{I. DFT computational details}

The first-principles calculations were conducted employing the density functional theory DFT+$U$ method~\cite{LDAU} with the Perdew-Burke- Ernzerhof (PBE) form of exchange correlation functional~\cite{PBE} and using the projector augmented plane wave basis-based method as implemented in the VASP~\cite{VASP1,VASP2}. We considered $U=$ 4.5 eV and $J_H=$ 0.95 eV as was used in the previous studies\cite{LFO113-HD,LFO-JM1,LFO-JM}. We considered Lu 4$f$ states in the core. Considering the polar $P6_3cm$ and non-polar $P\bar{3}c1$ phases structural relaxations were performed for both parent LuFeO$_3$ and the carrier doped systems employing 0.001 eV/\AA ~convergence criteria of the Hellmann-Feynman forces. In the present study, the effect of excess carrier was simulated by changing the number of electrons in the calculation and adding a homogeneous background charge to keep the system neutral~\cite{e-FE1,e-FE2}. We optimized the structures considering various magnetic orders. We used $6\times6\times2$ Monkhorst-Pack $\Gamma$ centred \textit{k}-point mesh and a kinetic energy cut-off value of 500 eV. We calculated the electric polarization using the Berry phase method~\cite{BP} as implemented in VASP.

In order to estimate the symmetric exchange interactions between the magnetic ions, we conducted total energy calculations of multiple collinear spin orders.  Additionally, the values of single-ion-anisotropy (SIA) parameters were calculated considering total energy of various non-collinear spin structures in the presence of spin-orbit ($L-S$) coupling as implemented in VASP~\cite{SO}.

\section{II. Improper ferroelectric phase}

\begin{figure*}
\begin{center}
\includegraphics[scale=0.5]{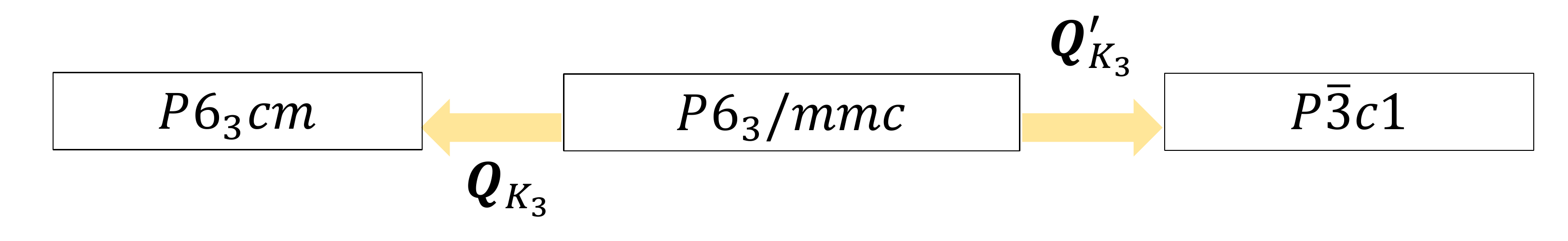} 
\end{center}
\caption{Phonon instability at the zone-boundary $\textbf{k}=(\frac{1}{3},\frac{1}{3},0)$ point with $K_3$ symmetry of the paraelectric $P6_3/mmc$ structure leads to polar $P6_3cm$ and non-polar $P\bar{3}c1$ phases. }
\label{FE0}
\end{figure*}

\begin{figure*}
\begin{center}
\includegraphics[scale=0.66]{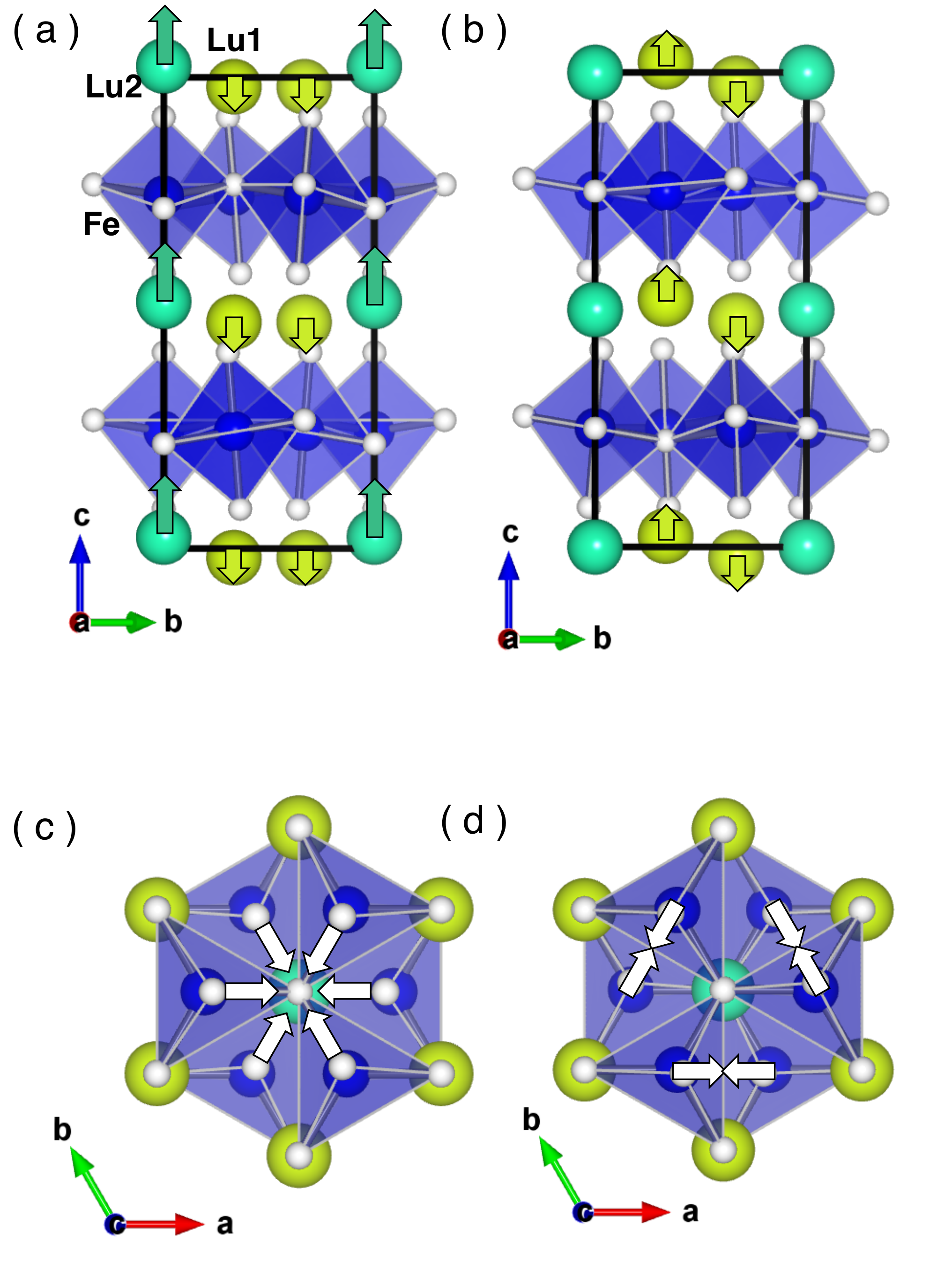} 
\end{center}
\caption{( a ) and (b) optimized polar $P6_3cm$ and non-polar $P\bar{3}c1$ structures illustrating respective Lu buckling displacements by arrows, respectively. (c) and (d) corresponding tilt pattern of the FeO$_5$ bipyramids.}
\label{FE1}
\end{figure*}

\begin{figure*}
\begin{center}
\includegraphics[scale=0.6]{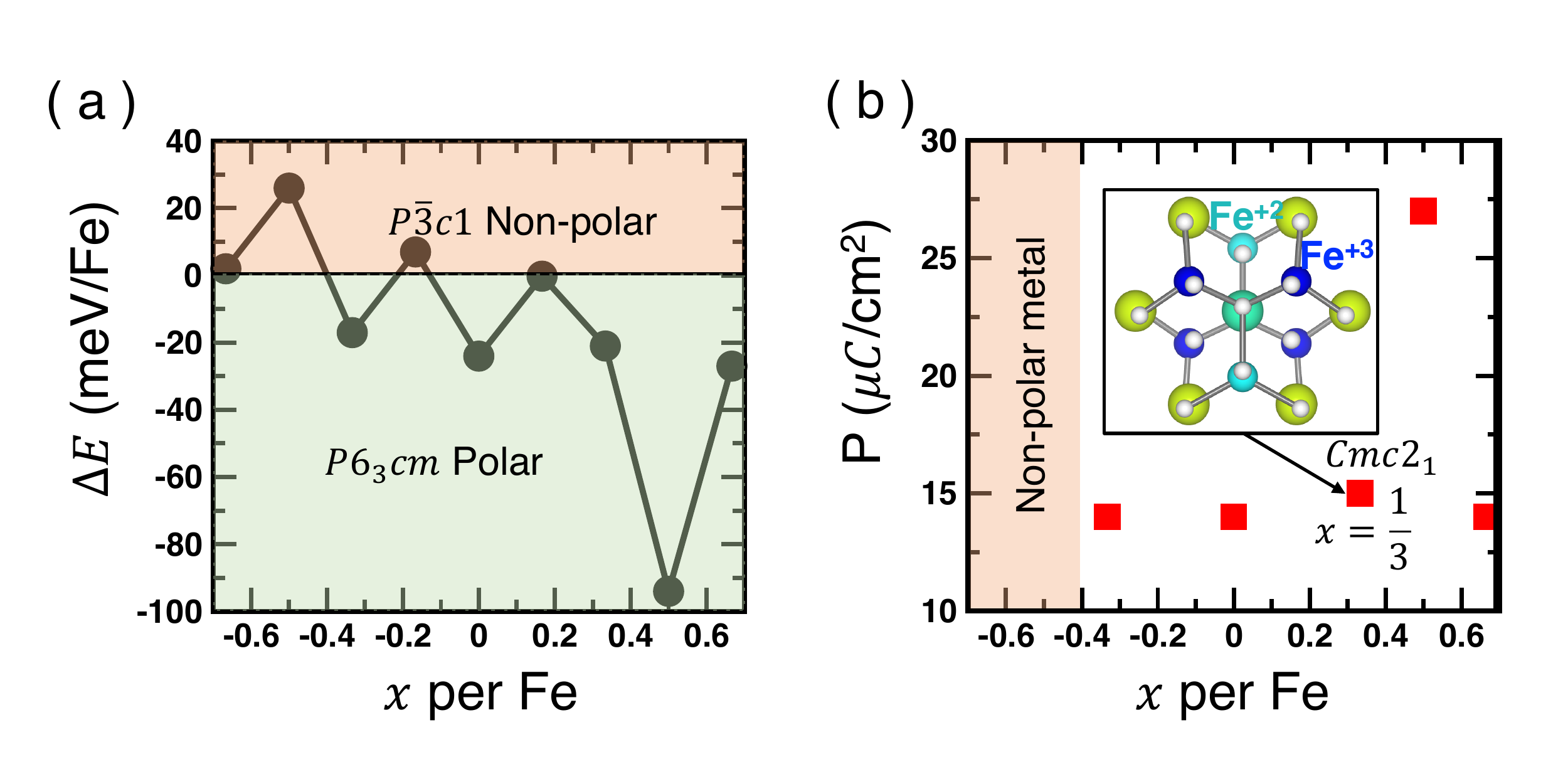} 
\end{center}
\caption{( a ) Relative stability of the polar phase with respect to the non-polar phase as a function of carrier doping concentrations $x$. ( b ) Calculated values of electric polarization directed along the crystallographic $c$ ($\hat{z}$) axis for the insulators. Inset shows the formation of Fe$^{2+}$ sublattice within Fe$^{3+}$ lattice.}
\label{FE}
\end{figure*}

The paraelectric phase of LuFeO$_3$ has $P6_3/mmc$ symmetry. The phonon instability with $K_3$ irreducible symmetry at the zone-boundary $\textbf{k}=(\frac{1}{3},\frac{1}{3},0)$ point can lead to either a polar $P6_3cm$ or a non-polar $P\bar{3}c1$ structure (see Fig.~\ref{FE0}). The corresponding structural distortion patterns are depicted in Fig.~\ref{FE1}.
Employing GGA+$U$ method we, therefore, optimized the polar and the non-polar structures by varying the concentration of carrier doping level, as shown in Fig.~\ref{FE}(a). We doped the system with both electrons and holes. As expected, the parent compound stabilizes in the $P6_3cm$ phase with an electric polarization of $\sim$ 14 $\mu$C/cm$^2$ directed along the crystallographic $\textbf{c}$ axis. 

Interestingly, while most of the electron doped systems are expected to crystallize in the FE insulating phase, the majority of the hole doped systems tends to crystallize in the non-polar metallic phase (see Fig.~\ref{FE}(a)). In particular, our results show formation of Fe$^{2+}$/Fe$^{3+}$ charged ordered (CO) C$_\textbf{q}$-type state, similar to LuFe$_2$O$_4$, for electron doping level of $\frac{1}{3}$ per Fe. In the C$_q$-type state, in each Fe layer, Fe$^{2+}$ triangular (T) lattice is situated within the Fe$^{3+}$ hexagonal lattice, as shown in Fig.~\ref{FE}(b). The doped system crystallizes in the $Cmc2_1$ structure exhibiting strong $\textbf{Q}^{\prime}_{K_3}$ distortions which induce a net polarization of $P \sim$ 15 $\mu$C/cm$^2$ slightly tilted from the crystallographic $\textbf{c}$ axis, as shown in Fig.~\ref{FE}(b). 

\section{III. Finite temperature Monte Carlo (MC) simulations}
\subsection{A. Computational details}

We performed classical Monte Carlo (MC) simulations using METROPOLIS algorithms~\cite{MC1,MC2} and proper periodic boundary conditions as implemented in our group MC package, to study magnetic phase transitions as a function of temperature and magnetic parameters. We calculated total energy $\xi(T)$ as a function of temperature ($T$) by considering $N_{MC}$ number of MC steps for each temperature step and performing $N_{Fe}$ spin-flips. Here $N_{Fe}$ represents total number of Fe ions, respectively, in the MC supercell structure. During each spin-flip process it randomly rotates the direction of the selected spin with a uniform probability distribution of the associated unit spin vector over a unit sphere. The specific heat as a function of temperature was calculated employing,
\begin{eqnarray}
C_v(T)=\dfrac{\langle \xi(T)^2\rangle - \langle \xi(T) \rangle^2}{k_B T^2}
\label{Cv}
\end{eqnarray}
where the angles bracket denotes thermal average. We also calculated the magnetization of the system as a function of temperature as,
\begin{equation}
M_{\varrho} = \frac{1}{N_{Fe}}\sum_{i=1}^{N_{Fe}}g\mu_BS_{\varrho}^i
\end{equation}
where, $\varrho = 1,2,3$ represent component of $\textbf{M}$ along the Cartesian axes $x$, $y$ and $z$, respectively. $\mu_B$ is the Bohr magneton and $g \approx 2$ denotes the gyromagnetic ratio.

\subsection{B. Model $H_1$}

\begin{figure*}
\begin{center}
\includegraphics[scale=0.5]{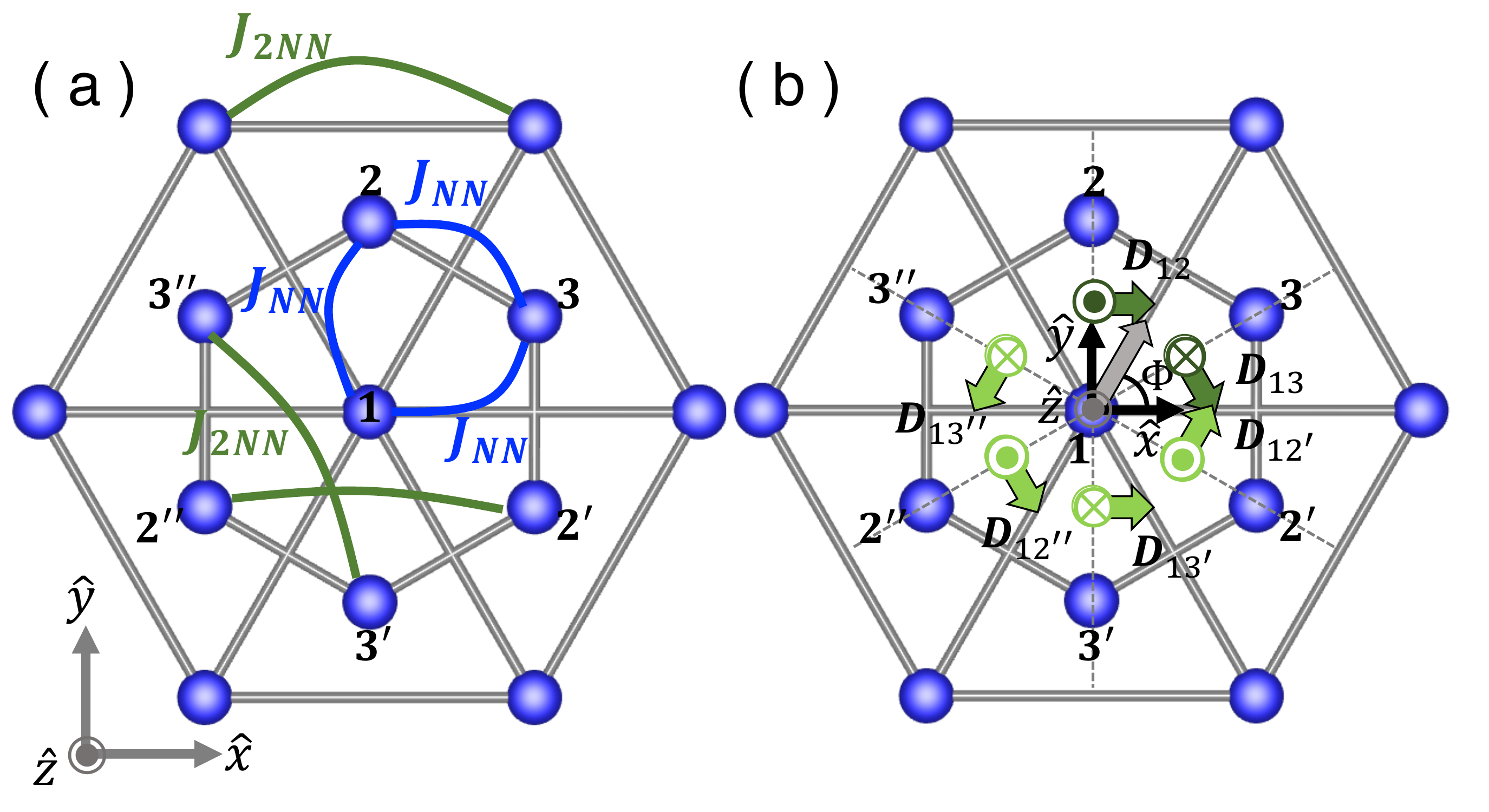} 
\end{center}
\caption{(a) Symmetric in-plane exchange interactions between Fe spins in the polar $P6_3cm$ structure. We considered $NN$ ($J_{NN}$) and $2NN$ ($J_{2NN}$) symmetric exchange (SE) interactions between Fe$^{3+}$ spins. (b) NN DM vectors acting on the Fe$^{3+}$ 1 for a single Fe layer. The longitudinal component of the DM vectors are shown with dot and cross marks representing respective orientation along the positive and negative $\hat{z}$ direction, respectively. The $\textbf{Q}_{K_3}$ distortions induce two non-equivalent transverse components of DM vectors mediated through non-equivalent planar oxygen O$_{p1}$ ($D_{12}= D_{13}=D^{[3+,3+]}$) and O$_{p2}$ ($D_{12^{\prime}} = D_{13^{\prime}} = D_{12^{\prime\prime}} = D_{13^{\prime\prime}} = D^{[3+,3+]\prime}$), respectively. DFT results indicate that~\cite{LFO113-HD}, the small difference in the magnitude of the DM vectors due to this non-equivalency can be ignored, i.e. $D^{[3+,3+]} = D^{[3+,3+]\prime}$. }
\label{LFO}
\end{figure*}

\begin{figure*}
\begin{center}
\includegraphics[scale=0.5]{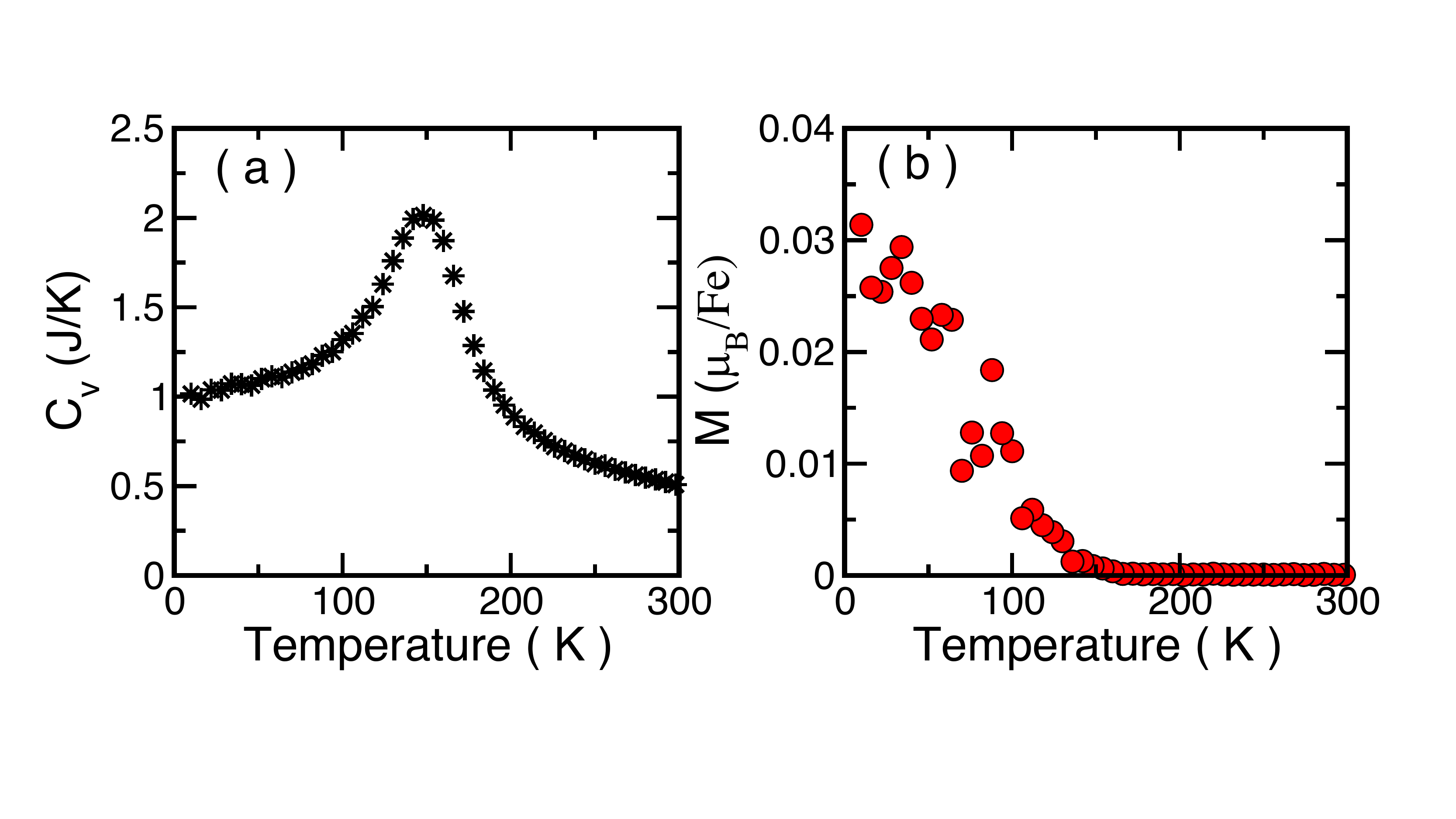} 
\end{center}
\caption{Calculated specific heat (a) and magnetization (b) as function of temperature considering $H_1$ model corresponding to the DFT estimated values of the magnetic parameters.}
\label{H1-MC1}
\end{figure*}

In the present study, we conducted MC simulations considering a spin model Hamiltonian of the parent system defined as,
\begin{equation}
H_{1}=H_{SE}+H_{DM}+H_{SIA}
\end{equation}
where the first term denotes the energy contribution due to the symmetric exchange (SE) interactions between Fe$^{3+}$ spins and given by,
\begin{equation}
\begin{split}
H_{SE}=\sum_{\langle i,j \rangle_{NN}}J_{NN}\textbf{S}^{3+}_{i} \cdot \textbf{S}^{3+}_{j}+\sum_{\langle i,j \rangle_{2NN}}J_{2NN}\textbf{S}^{3+}_{i} \cdot \textbf{S}^{3+}_{j}\\+\sum_{\langle i,j \rangle_{c}}J_{c}^{ij}\textbf{S}^{3+}_{i} \cdot \textbf{S}^{3+}_{j}
\end{split}
\end{equation}
it consists of nearest-neighbor (NN) and second-nearest-neighbor (2NN) in-plane SE interactions between Fe$^{3+}$ ($S^{3+} = \frac{5}{2}$) spins (see Fig.~\ref{LFO}(a)). In the parent system, we considered that the Fe spins interact via six almost equivalent NN and six equivalent 2NN connecting pathways, as the difference in the strength of NN SE interactions mediated through non-equivalent planar oxygens (O$_{p1}$ and O$_{p2}$) is expected to be negligible. $i$ and $j$ denote the site of the Fe$^{3+}$ spins and each pair was counted once. For the parent system, let us denote,
\begin{equation}
\begin{split}
J_{NN}=J_{NN}^{[3+,3+]}\\
J_{2NN}=J_{2NN}^{[3+,3+]}
\end{split}
\end{equation}
In addition, SE interactions between the consecutive Fe layers mediated via Fe-O-Lu-O-Fe pathways, $J_c^{ij}$, were also taken into account. The $\textbf{Q}_{K_3}$ distortion splits six equivalent inter-layer SE interactions into two and four non-equivalent SE interactions, namely $J_c$ and $J^{\prime}_c$, respectively. The important parameter that contribute to stabilize a magnetic order in this structure is the effective inter-layer SE interaction defined as, $\Delta J^{[3+,3+]}_c = J_c - J^{\prime}_c$.
\begin{table*}
\begin{center}
\caption{GGA+$U$ estimated values of SE interactions between Fe spins considering $U = $ 4.5 eV and $J_H =$ 0.95 eV for both parent and $e^{-}$ doped system. $^{*}$ denotes the values reported in Ref.~\cite{LFO113-HD}.}
\begin{tabular}{c|ccc|c|cccc}
\hline
\hline
&$J_{NN}^{[3+,3+]}$&$J_{2NN}^{[3+,3+]}$&$\Delta J_{c}^{[3+,3+]}$&&$J_{NN}^{[3+,3+]}$&$J_{2NN}^{[3+,3+]}$&$J_{NN}^{[2+,3+]}$&$J_{2NN}^{[2+,2+]}$\\
LuFeO$_3$&(meV)&(meV)&(meV)&$e^-$- LuFeO$_3$&(meV)&(meV)&(meV)&(meV)\\
&&&&&&&&\\
&6.3 (6.3$^{*}$)&0.3 (0.5$^{*}$)&0.4 (0.38$^{*}$)&&8.5&0.3&1.1&-0.1\\
\hline
\hline
\end{tabular}
\label{J}
\end{center}
\end{table*}
In order to estimate the values of these parameters for the parent system we calculated total energy of various spin configurations using a $2\times2\times1$ supercell of the $P6_3cm$ structure. The estimated values are tabulated in Table~\ref{J}. 

In the paraelectric (PE) $P6_3/mmc$ phase, only the NN DM interactions parallel to the $\hat{z}$ axis are allowed and the symmetry adopted patterns are shown in 
Fig.~\ref{LFO}(b) around the ion 1. The $\textbf{Q}_{K_3}$ distortions 
induce the in-plane components, where around ion 1, two and four of the NN 
DM vectors form clockwise and  anti-clockwise rotational patterns mediated via planar oxygen O$_{p1}$ and O$_{p2}$, respectively. 
(see Fig.~\ref{LFO}(b)). Accordingly, let us define, 
$D_{12} = D_{13} = D_{23} = D^{[3+,3+]}$ and $D_{12^{\prime}} = 
D_{12^{\prime\prime}} = D_{13^{\prime}} = D_{13^{\prime\prime}} = 
D_{23^{\prime}} = D_{23^{\prime}} = D_{32^{\prime}} = 
D_{32^{\prime\prime}} = D^{[3+,3+]\prime}$. However, 
previous DFT calculations indicate that $D^{[3+,3+]} \sim D^{[3+,3+]\prime}$~\cite{LFO113-HD}. Therefore, in order to simplify the spin model, we have used this postulate throughout the present study. The corresponding energy contribution of a single Fe layer ($L_1$) shown in Fig.~\ref{LFO}(b) is given by,
\begin{equation}
\begin{split}
E_{DM}^{L_1}|_{H_1} = \bar{\textbf{D}}_{12}|_{L_1}\cdot \textbf{S}^{3+}_1 \times \textbf{S}^{3+}_2 + \bar{\textbf{D}}_{13}|_{L_1}\cdot \textbf{S}^{3+}_1 \times \textbf{S}^{3+}_3 \\+ \bar{\textbf{D}}_{23}|_{L_1}\cdot \textbf{S}^{3+}_2 \times \textbf{S}^{3+}_3  
\end{split}
\end{equation}
where the effective DM vectors can be defined as,
\begin{eqnarray}
\bar{\textbf{D}}_{12}|_{L1} = \textbf{D}_{12}+\textbf{D}_{12^{\prime}} + \textbf{D}_{12^{\prime\prime}}=\bar{\textbf{D}}_{xy}(\varphi)+\bar{\textbf{D}}_{z}\\ 
\bar{\textbf{D}}_{13}|_{L1} = \textbf{D}_{13}+\textbf{D}_{13^{\prime}} + \textbf{D}_{13^{\prime\prime}}=\bar{\textbf{D}}_{xy}(\varphi+\frac{5\pi}{3})-\bar{\textbf{D}}_{z}\\
\bar{\textbf{D}}_{23}|_{L1} = \textbf{D}_{23}+\textbf{D}_{23^{\prime}} + \textbf{D}_{23^{\prime\prime}}=\bar{\textbf{D}}_{xy}(\varphi+\frac{4\pi}{3})+\bar{\textbf{D}}_{z} 
\end{eqnarray}
Where $\varphi$ represent the angle between the transverse component of the effective $\bar{\textbf{D}}_{12}|_{L1}$ with $\hat{x}$ axis. Also, $\bar{D}_{xy}= 2D_{xy}^{[3+,3+]}$ and $\bar{D}_{z}= 3D_{z}^{[3+,3+]}$.
As the consecutive Fe triangular layers, are connected through a $\tilde{2}_c$ axis, the associated effective transverse components of the DM vectors are anti-parallel to each other. 

The SIA tensor of the Fe$^{3+}$ ions consists of non-equivalent diagonal components ($\tau_{xx} \neq \tau_{yy} \neq \tau_{zz}$) and off-diagonal components  $\tau_{xz}=\tau_{zx}$. Considering the zero trace condition, $\tau_{zz}=-(\tau_{xx}+\tau_{yy})$. In the PE $P6_3/mmc$ phase $\tau_{xx} = \tau_{yy}$ and $\tau_{xz}=\tau_{zx}=0$. Our present results and previous reports~\cite{LFO113-HD} suggest, $\tau_{xx} \sim \tau_{yy}=\tau^{3+}$. Additionally, to further simplify the model we assumed $\tau_{xz}\rightarrow0$, as the effect of this off-diagonal component will be captured by the transverse component of the DM interactions. We therefore employed a simplified form of SIA tensor as,
\begin{equation}
\hat{\tau}^{3+} = 
\begin{pmatrix}
\tau^{3+} & 0 & 0\\
0 & \tau^{3+} & 0\\
0&0&-2\tau^{3+}
\end{pmatrix}
\end{equation}
The positive and negative value of $\tau^{3+}$ indicates uni-axial ($\hat{z}$) and uni-planar ($xy$) magnetic anisotropy, respectively. The DFT estimated values of SIA parameters are listed in Table~\ref{SIA1}.

\begin{table*}
\begin{center}
\caption{GGA+$U$ estimated values of SIA parameters and NN DM interactions between magnetic ions considering $U = $ 4.5 eV and $J_H =$ 0.95 eV for both parent and $e^{-}$ doped system. $^{*}$ denotes the values reported in Ref.~\cite{LFO113-HD}.}
\begin{tabular}{ccc|ccc}
\hline
\hline
&LuFeO$_3$&&&$e^-$- LuFeO$_3$&\\
$\tau^{3+}$&$D_{xy}^{[3+,3+]}$&$D_{z}^{[3+,3+]}$&&$\tau^{3+}$&$\tau^{2+}$\\
(meV)&(meV)&(meV)&&(meV)&(meV)\\
\hline
0.03 (0.08$^{*}$) &0.1$^{*}$&0.06$^{*}$&&0.03&0.21\\
\hline
\hline
\end{tabular}
\label{SIA1}
\end{center}
\end{table*}

We report the results of Monte Carlo (MC) simulations performed on an 8$\times$8$\times$6 cell consisting of 2304 magnetic ions with spin value $S^{3+}=\frac{5}{2}$, and considering 10$^9$ MC steps for each temperature using the model Hamiltonian $H_1$. The convergence of $\xi(T)$ and the ground state magnetic structure were cross checked by considering upto 10$\times$10$\times$10 ($N_{ion}=$ 6000) cell size and 10$^9$ MC steps. We considered both cooling and heating processes to carefully cross check the stability of the magnetic phases. During these processes, the final simulated magnetic configuration corresponding to a particular temperature was considered as the initial magnetic configuration corresponding to the next value of temperature. We conducted finite temperature MC simulations considering a wide magnetic parameter space of DM interactions between the magnetic ions and as well as their SIA. 

\begin{figure*}
\begin{center}
\includegraphics[scale=0.5]{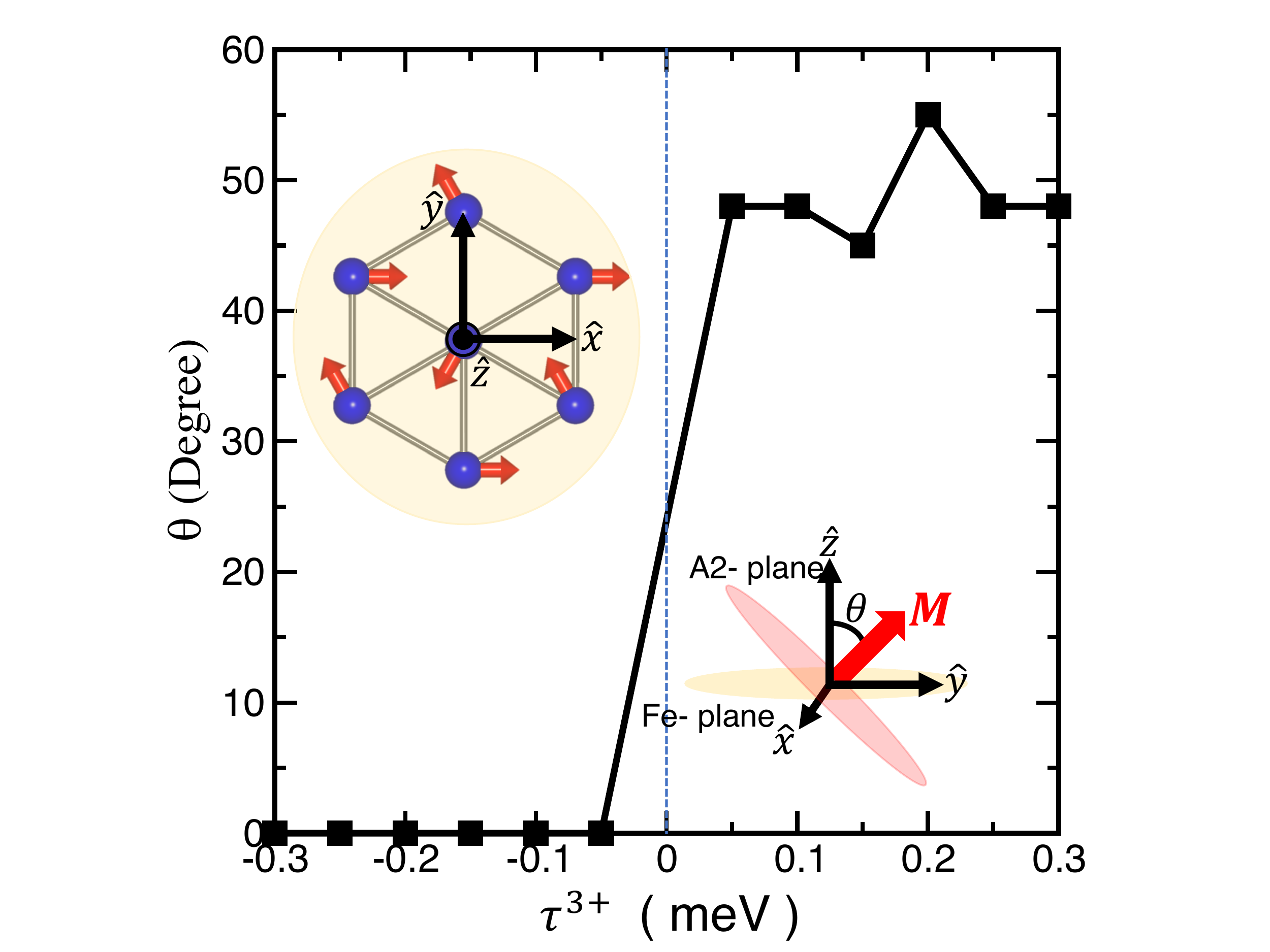} 
\end{center}
\caption{The calculated tilt angle $\theta$ of the A2 spin order plane as a function of $\tau^{3+}$.}
\label{H1-MC2}
\end{figure*}

\subsection{C. Model $H_2$}

\begin{figure*}
\begin{center}
\includegraphics[scale=0.6]{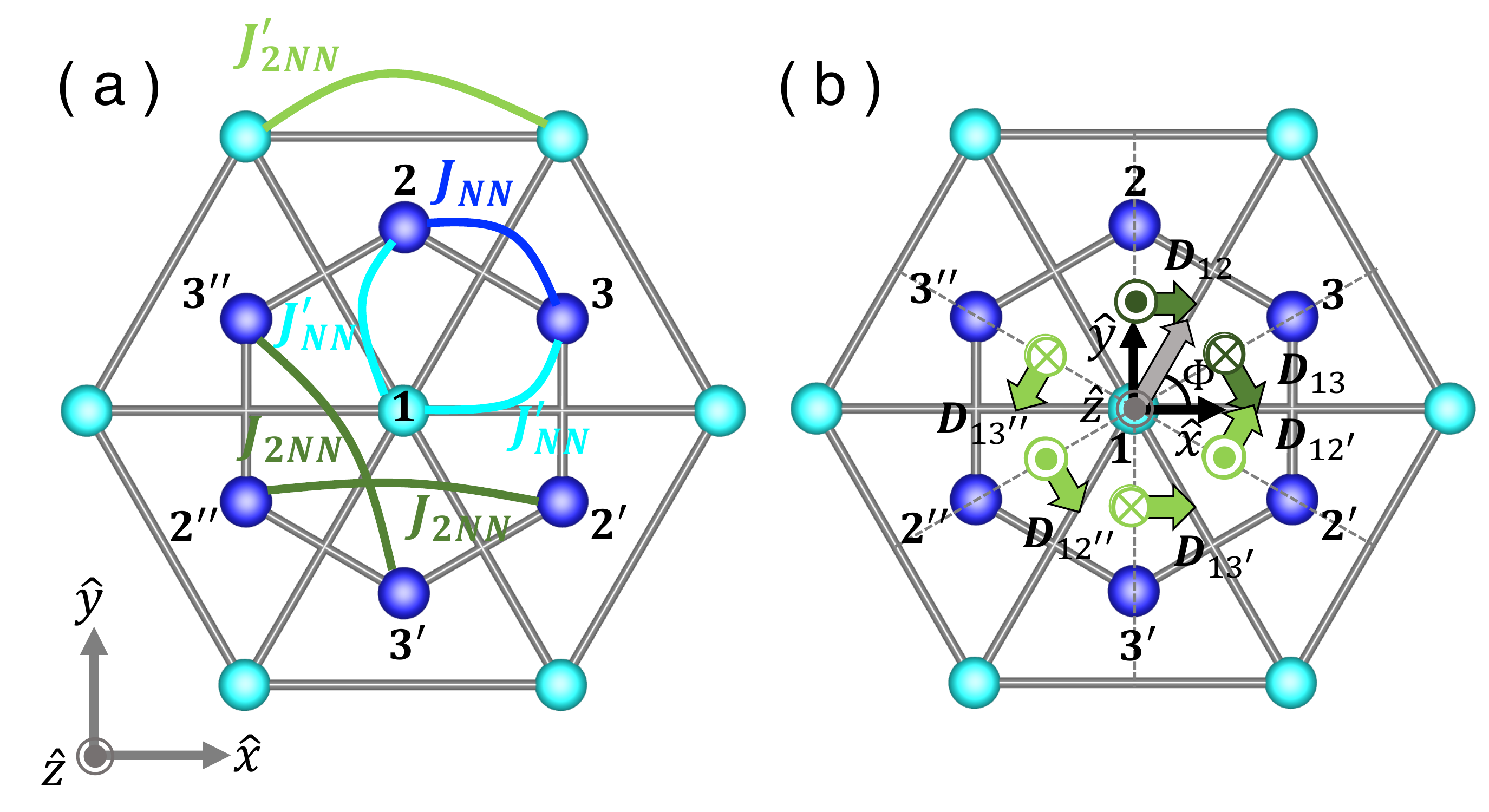} 
\end{center}
\caption{(a) Symmetric in-plane exchange interactions between Fe spins in the polar $Cmc2_1$ structure, respectively. We considered inter- ($J_{NN}^{\prime}$) and intra- ($J_{NN}$) sublattice $NN$ symmetric exchange (SE) interactions. Also, the intra-sublattice $2NN$ SE interactions ($J_{2NN}$ and $J_{2NN}^{\prime}$) were taken into account. (b) NN DM vectors acting on the Fe$^{2+}$ 1 for a single Fe layer. The longitudinal component of the DM vectors are shown with dot and cross marks representing respective orientation along the positive and negative $\hat{z}$ direction, respectively. We assume the same $NN$ DM vector pattern in the $e^-$ doped structure as it is in the parent system, because $\textbf{Q}_{K_3}$ remains as the primary order parameter of its FE behavior. Similar to the parent system, the $\textbf{Q}_{K_3}$ distortions induce two non-equivalent transverse components of DM vectors mediated through non-equivalent planar oxygen O$_{p1}$ ($D_{12}= D_{13}=D^{[2+,3+]}$) and O$_{p2}$ ($D_{12^{\prime}} = D_{13^{\prime}} = D_{12^{\prime\prime}} = D_{13^{\prime\prime}} = D^{[2+,3+]\prime}$), respectively. However, the small difference in the magnitude of the DM vectors due to this non-equivalency can be ignored to simplify the model, i.e. $D^{[2+,3+]} = D^{[2+,3+]\prime}$. Similarly, one can describe $NN$ Fe$^{3+}$-Fe$^{3+}$ DM vectors with magnitude $D^{[3+,3+]}$.}
\label{ELFO}
\end{figure*}

\begin{figure*}
\begin{center}
\includegraphics[scale=0.6]{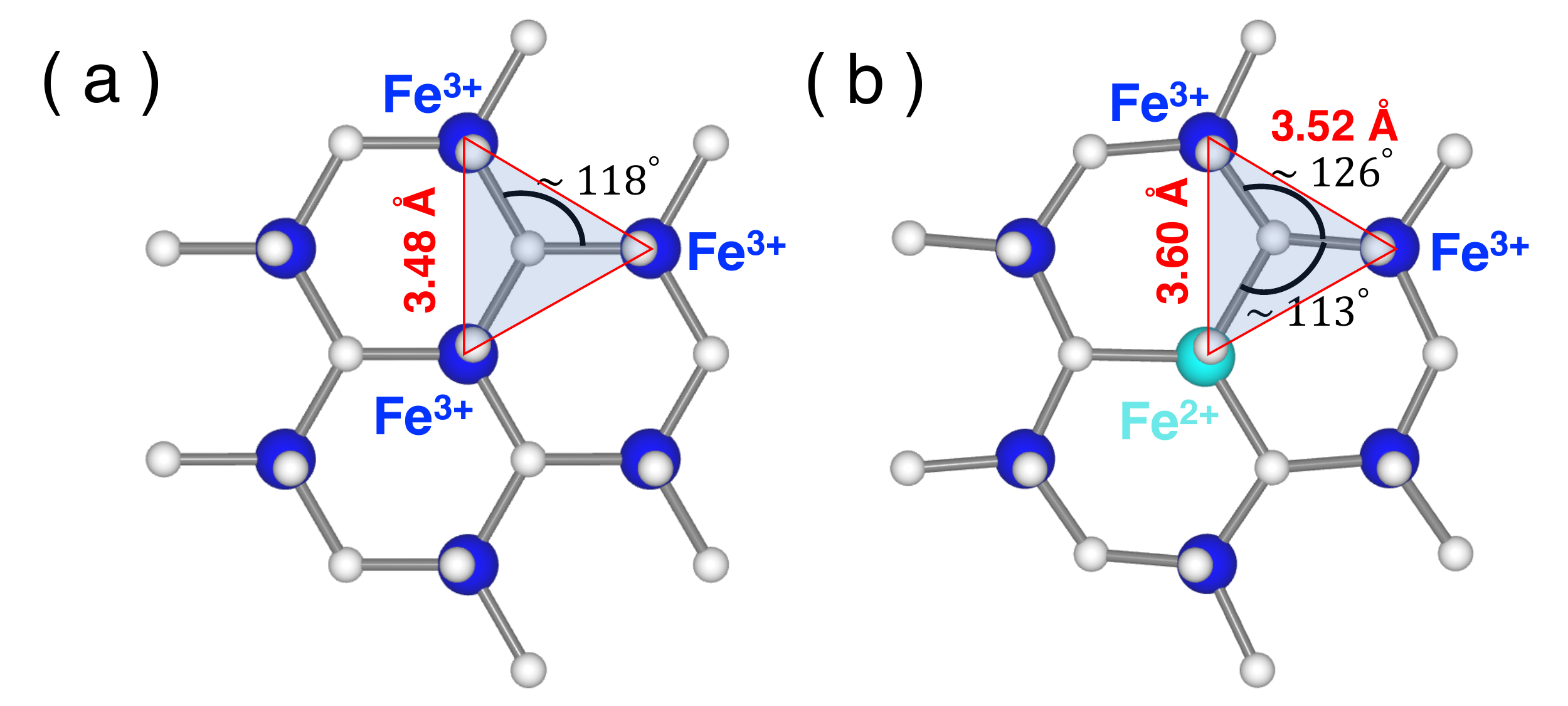} 
\end{center}
\caption{Symmetric NN in-plane exchange interaction pathways between Fe spins in the polar parent (a) LuFeO$_3$ and (b) $e^{-}$ - doped system. The Fe$^{3+}$ and Fe$^{2+}$ ions are shown with solid blue and cyan spheres, respectively.}
\label{PATH}
\end{figure*}

The electron doped system shows a transition from $P6_3cm$ $\rightarrow$ $Cmc2_1$ symmetry due the formation of C$_\textbf{q}$ charge order. It introduces two magnetic sublattices, such as Fe$^{2+}$ ($S^{2+}$) triangular ($T$) and Fe$^{3+}$ ($S^{3+}$)  hexagonal ($H$). We incorporated the changes in the spin Hamiltonian accordingly,

\begin{equation}
H_{2}=H^{\prime}_{SE}+H^{\prime}_{DM}+H^{\prime}_{SIA}
\end{equation}
where the first term denotes SE interactions between magnetic ions and is given by,
\begin{equation}
\begin{split}
H^{\prime}_{SE}=\sum_{\langle i_H,j_H \rangle_{NN}}J_{NN}\textbf{S}^{3+}_{i_H} \cdot \textbf{S}^{3+}_{j_H}+\sum_{\langle i_T,j_H \rangle_{NN}}J^{\prime}_{NN}\textbf{S}^{2+}_{i_T} \cdot \textbf{S}^{3+}_{j_H}\\+\sum_{\langle i_T,j_T \rangle_{2NN}}J^{\prime}_{2NN}\textbf{S}^{2+}_{i_T} \cdot \textbf{S}^{2+}_{j_T}+\sum_{\langle i_H,j_H \rangle_{2NN}}J_{2NN}\textbf{S}^{3+}_{i_H} \cdot \textbf{S}^{3+}_{j_H}\\+\sum_{\langle i_T,j_T \rangle_{c}}J_{c}^{i_Tj_T}\textbf{S}^{2+}_{i_T} \cdot \textbf{S}^{2+}_{j_T} + \sum_{\langle i_T,j_H \rangle_{c}}J_{c}^{i_Tj_H}\textbf{S}^{2+}_{i_T} \cdot \textbf{S}^{3+}_{j_H} \\+ \sum_{\langle i_H,j_H \rangle_{c}}J_{c}^{i_Hj_H}\textbf{S}^{3+}_{i_H} \cdot \textbf{S}^{3+}_{j_H}
\end{split}
\end{equation}
where $(i_T,j_T)$ and $(i_H,j_H)$ denote the site of the Fe$^{2+}$ ($S^{2+}=2$) and Fe$^{3+}$ ($S^{3+}=\frac{5}{2}$) spins in the $T$ and $H$ sublattices (see Fig.~\ref{ELFO}(a)), respectively.
Similar to the parent system, we neglected the slight differences in the strength of the NN SE interactions and considered each Fe$^{2+}$ spin interacts with NN Fe$^{3+}$ spins mediated via six equivalent pathways, \textit{i.e. },
\begin{equation}
J_{NN}^{\prime}=J_{NN}^{[2+,3+]}
\end{equation}
and with six equivalent 2NN Fe$^{2+}$ spins as,
\begin{equation}
J_{2NN}^{\prime}=J_{2NN}^{[2+,3+]}
\end{equation}
On the other hand, the intra-Fe$^{3+}$ sublattice interactions are described as,
\begin{equation}
\begin{split}
J_{NN}=J_{NN}^{[3+,3+]}\\
J_{2NN}=J_{2NN}^{[3+,3+]}
\end{split}
\end{equation}
The estimated values of SE interactions are given in Table~\ref{J}.
The antiferromagnetic (AFM) NN interaction between Fe$^{3+}$ spins increases under the influence of the doped electrons to a value of 8.5 eV. This behavior can be associated with the increase in the superexchange $\angle$Fe-O-Fe pathways from $118^{\circ} \rightarrow 126^{\circ}$  (see Fig.~\ref{PATH}). Note that, $180^{\circ}$ and $90^{\circ}$ superexchange connecting angles prefer AFM and ferromagnetic (FM) interaction, respectively. The $J_{NN}^{[2+,3+]}$ is also AFM in nature, but almost an order of magnitude smaller than $J_{NN}^{[3+,3+]}$. This can be attributed to various factors, such as lowering of the $\angle$Fe$^{2+}$-O-Fe$^{3+}$ (see Fig.~\ref{PATH}) of the mediating pathways and FM components due to the Fe$^{2+}$-Fe$^{3+}$ multi-orbital hopping process. While $2NN$ within Fe$^{3+}$ sublattice remains AFM in nature and weak, the $2NN$ interaction between Fe$^{2+}$ spins was found to be FM and weak in nature. 

The inter-layer SE interaction is complex in nature. Each Fe$^{3+}$ ions experience four Fe$^{3+}$-Fe$^{3+}$ and two Fe$^{2+}$-Fe$^{3+}$ interactions mediated through long Fe-O-Lu-O-Fe pathways with an effective interaction $\Delta J_c^{3+}$. On the other hand, each Fe$^{2+}$ ions experience four Fe$^{2+}$-Fe$^{3+}$ and two Fe$^{2+}$-Fe$^{2+}$ interactions mediated through long Fe-O-Lu-O-Fe pathways with an effective interaction $\Delta J_c^{2+}$. While we consider an effective $\Delta J_c^{3+}$ of same strength and nature as of the parent system, \textit{i.e.} $\sim$ 0.4 meV, we ignored $\Delta J_c^{2+}$ in the present MC simulations.

In the $e^{-}$ doped $Cmc2_1$ structure considering the NN DM vector pattern same as the parent system, as depicted in ~\ref{ELFO}(b), the energy contribution of a single Fe layer ($L_1$) is given by,
\begin{equation}
\begin{split}
E_{DM}^{L_1}|_{H_2} = \bar{\textbf{D}}^{[2+,3+]}_{12}|_{L_1}\cdot \textbf{S}^{2+}_1 \times \textbf{S}^{3+}_2 \\+ \bar{\textbf{D}}^{[2+,3+]}_{13}|_{L_1}\cdot \textbf{S}^{2+}_1 \times \textbf{S}^{3+}_3 \\+ \bar{\textbf{D}}^{[3+,3+]}_{23}|_{L_1}\cdot \textbf{S}^{3+}_2 \times \textbf{S}^{3+}_3  
\end{split}
\end{equation}
Similar to the parent system we ignored the small non-equivalency effects. 
The effective DM vectors can be defined as,
\begin{equation}
\begin{split}
\bar{\textbf{D}}^{[2+,3+]}_{12}|_{L1} = \textbf{D}_{12}+\textbf{D}_{12^{\prime}} + \textbf{D}_{12^{\prime\prime}}\\=\bar{\textbf{D}}^{[2+3+]}_{xy}(\varphi)+\bar{\textbf{D}}^{[2+,3+]}_{z}
\end{split}
\end{equation}
\begin{equation}
\begin{split}
\bar{\textbf{D}}^{[2+,3+]}_{13}|_{L1} = \textbf{D}_{13}+\textbf{D}_{13^{\prime}} + \textbf{D}_{13^{\prime\prime}}\\=\bar{\textbf{D}}^{[2+,3+]}_{xy}(\varphi+\frac{5\pi}{3})-\bar{\textbf{D}}^{[2+,3+]}_{z}
\end{split}
\end{equation}
\begin{equation}
\begin{split}
\bar{\textbf{D}}^{[3+,3+]}_{23}|_{L1} = \textbf{D}_{23}+\textbf{D}_{23^{\prime}} + \textbf{D}_{23^{\prime\prime}}\\=\bar{\textbf{D}}^{[3+,3+]}_{xy}(\varphi+\Delta\varphi+\frac{4\pi}{3})+\bar{\textbf{D}}^{[3+,3+]}_{z} 
\end{split}
\end{equation}
Where $\varphi$ represent the angle between the transverse component of the effective $\bar{\textbf{D}}^{[2+,3+]}_{12}|_{L1}$ with $\hat{x}$ axis. For clarity we assumed $\Delta\varphi = 0$. Also, $\bar{D}^{[2+,3+]}_{xy}= 2D_{xy}^{[2+,3+]}$,  $\bar{D}^{[3+,3+]}_{xy}= 2D_{xy}^{[3+,3+]}$, $\bar{D}^{[2+,3+]}_{z}= 3D_{z}^{[2+,3+]}$ and $\bar{D}^{[3+,3+]}_{z}= 3D_{z}^{[3+,3+]}$.
As the consecutive Fe triangular layers are connected through a $\tilde{2}_c$ axis, the associated effective transverse components of the DM vectors are anti-parallel to each other. 

Assuming similar postulates as developed in the case of the parent system, we therefore employed a simplified form of SIA tensor as,
\begin{equation}
\hat{\tau}^{3+} = 
\begin{pmatrix}
\tau^{3+} & 0 & 0\\
0 & \tau^{3+} & 0\\
0&0&-2\tau^{3+}
\end{pmatrix}
\end{equation}

\begin{equation}
\hat{\tau}^{2+} = 
\begin{pmatrix}
\tau^{2+} & 0 & 0\\
0 & \tau^{2+} & 0\\
0&0&-2\tau^{2+}
\end{pmatrix}
\end{equation}

The positive and negative value of $\tau^{3+}$ ($\tau^{2+}$) indicates uni-axial ($\hat{z}$) and uni-planar ($xy$) magnetic anisotropy of the Fe$^{3+}$ (Fe$^{2+}$) ions, respectively. The DFT estimated values of SIA parameters are given in Table~\ref{SIA1}.

Similar to $H_1$ model, we conducted MC simulations performed on the $H_2$ model considering an 8$\times$8$\times$6 cell consisting of 1536 Fe$^{3+}$ ($S^{3+}=\frac{5}{2}$) and 768 Fe$^{2+}$ ($S^{2+}=2$) magnetic ions, and considering 10$^9$ MC steps for each temperature using the $H_2$ model constructed through GGA+$U$  calculations. The convergence of $\xi(T)$ and the ground state magnetic structure were cross checked by considering upto 10$\times$10$\times$10 ($N_{ion}=$ 6000) cell size and 10$^9$ MC steps. We primarily conducted finite temperature MC simulations considering $\textbf{D}^{[2+,3+]}$ magnetic parameter space using the values of $\textbf{D}^{[3+,3+]}$ as estimated in the parent system and  the estimated SIA parameters and SE interactions. This approach allowed us to explore the effect of $\textbf{D}^{[2+,3+]}$ on the stability of the magnetic order, as well as to develop the coupling between the magnetic order and the improper FE distortion $\textbf{Q}_{K_3}$, hence with electric polarization $\textbf{P}$.

\section{IV. Ferrimagnetic order}

\begin{figure*}
\begin{center}
\includegraphics[scale=0.6]{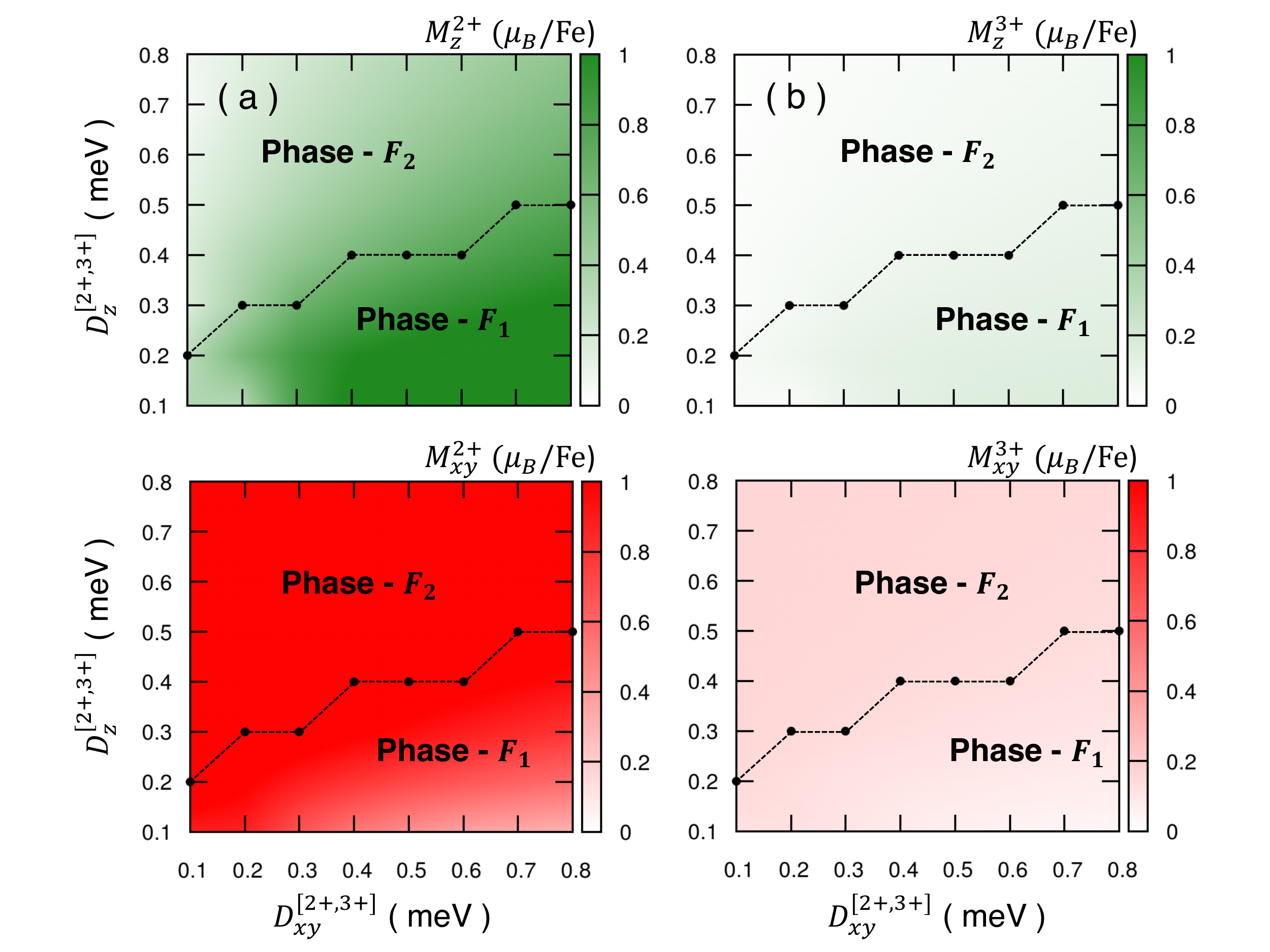} 
\end{center}
\caption{Calculated sublattice magnetization using $H_2$ model as functions of the magnitude of the transverse and longitudinal components of $\textbf{D}^{[2+,3+]}$}
\label{S-MC2}
\end{figure*}

Finite temperature MC simulations predicted three ferrimagnetic phases. Two magnetic order parameters can be defined as, (1) net magnetization $\textbf{M}$ of the $T$ sublattice and (2) AFM $\textbf{L}^{3+}$ order in the H sublattice. As the magnetization is primarily contributed by the Fe$^{2+}$ sublattice, as shown in Fig.~\ref{S-MC2}, one can define the phase $F_1$ as,
\begin{eqnarray}
\textbf{M}_1^{2+}=\textbf{M}=\Delta \textbf{M}_{xy}+\textbf{M}_{z}\\
\textbf{M}_2^{3+}=\textbf{M}_{xy}^{3+}; \textbf{M}_3^{3+}=-\textbf{M}_{xy}^{3+}; \textbf{L}^{3+}=\textbf{M}_2^{3+}-\textbf{M}_3^{3+}
\end{eqnarray}
The energy contribution due to Fe$^{2+}$ - Fe$^{3+}$ DM interaction is given by,
\begin{equation}
E_{DM}^{F_1}=\textbf{M}_z \cdot \Delta \textbf{D}_{xy}^{[2+,3+]} \times \textbf{L}^{3+} + \Delta \textbf{M}_{xy} \cdot \Delta \textbf{D}_{z}^{[2+,3+]} \times \textbf{L}^{3+}
\end{equation}
where the effective DM interactions are defined as, $\Delta \textbf{D}_{xy}^{[2+,3+]} = 2(\bar{\textbf{D}}^{[2+,3+]}_{xy}(\varphi+\frac{5\pi}{3})-\bar{\textbf{D}}^{[2+,3+]}_{xy}(\varphi)) \sim (2D_{xy}^{[2+,3+]}cos\Phi,2D_{xy}^{[2+,3+]}sin\Phi,0)$ and $\Delta \textbf{D}_{z}^{[2+,3+]}=-4\bar{\textbf{D}}^{[2+,3+]}_{z}$. One can derive,
\begin{eqnarray}
\textbf{M}_z \propto \Delta \textbf{D}_{xy}^{[2+,3+]} \times \textbf{L}^{3+} \propto \textbf{Q}_{K_3} \times \textbf{L}^{3+}\\
\Delta \textbf{M}_{xy} \propto \Delta \textbf{D}_{z}^{[2+,3+]} \times \textbf{L}^{3+}
\end{eqnarray}

\begin{figure*}
\begin{center}
\includegraphics[scale=0.5]{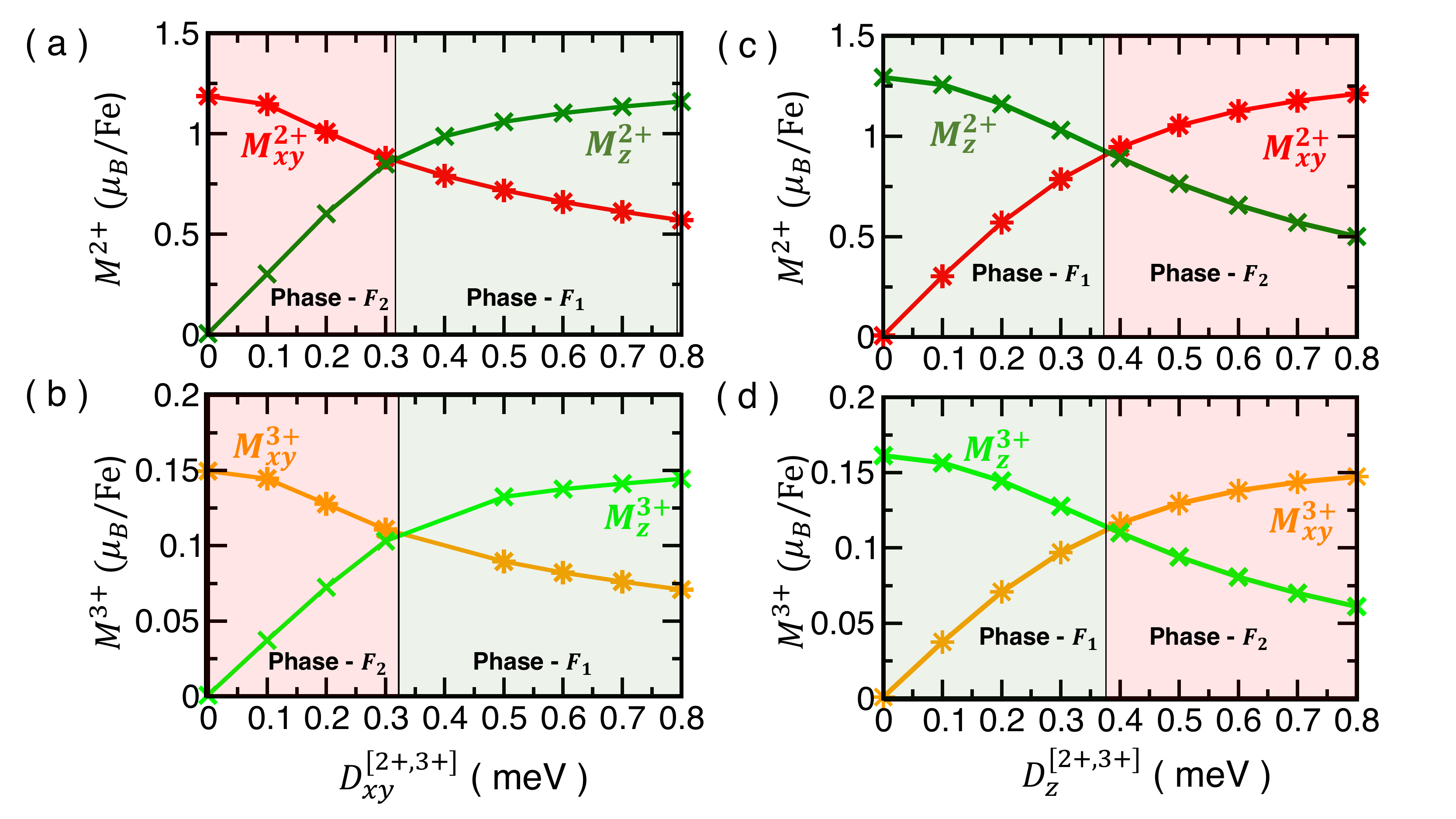} 
\end{center}
\caption{(a) and (b) Calculated sublattice magnetization using $H_2$ model as a function of the magnitude of the transverse component of $\textbf{D}^{[2+,3+]}$ considering a fixed value of $D^{[2+,3+]_{z}}=$ 0.3 meV. (c) and (d) Calculated sublattice magnetization using $H_2$ model as a function of the magnitude of the longitudinal component of $\textbf{D}^{[2+,3+]}$ considering a fixed value of $D^{[2+,3+]_{xy}}=$ 0.3 meV.}
\label{S-MC3}
\end{figure*}

\begin{figure*}
\begin{center}
\includegraphics[scale=0.6]{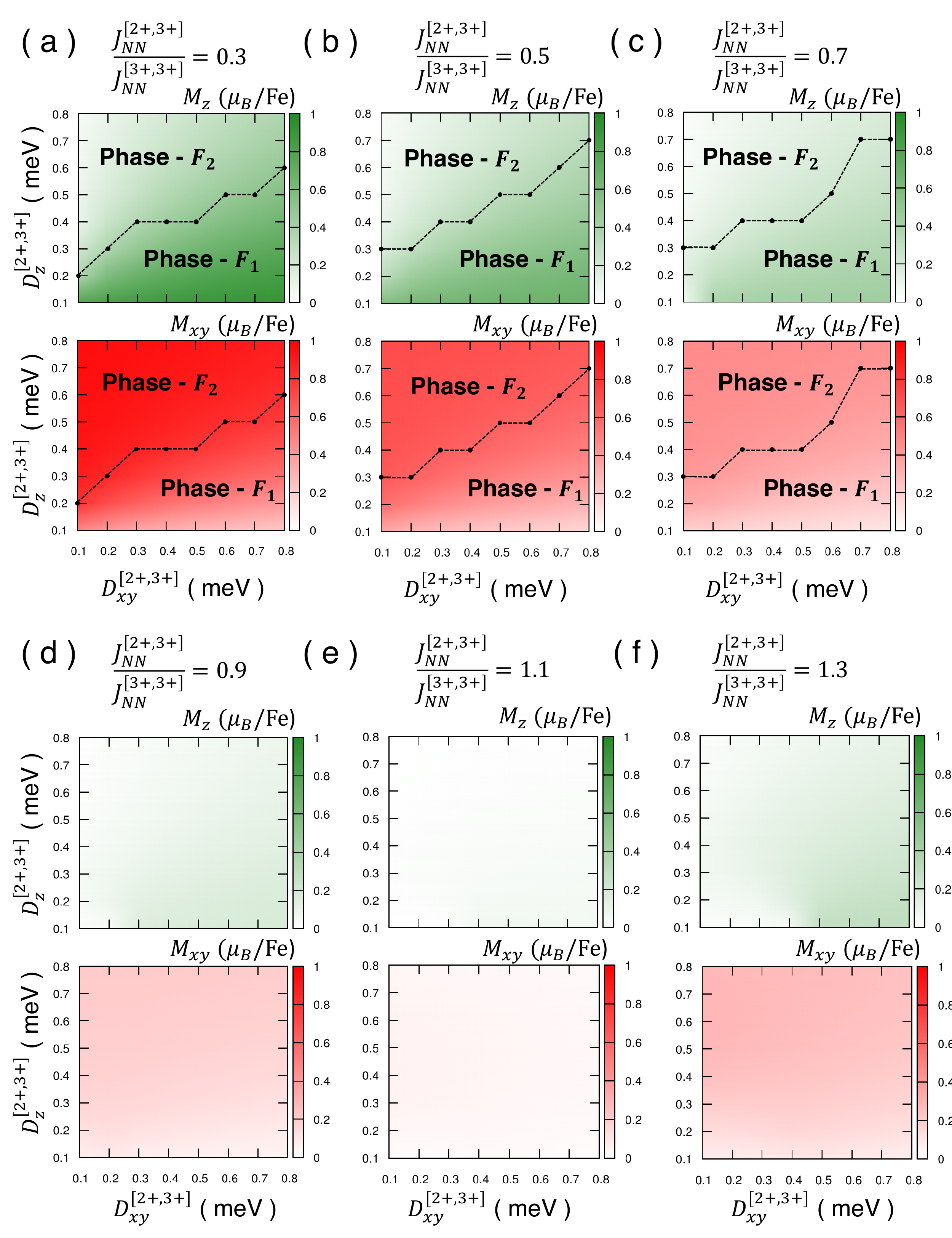} 
\end{center}
\caption{Calculated net magnetization using $H_2$ model as functions of the magnitude of the transverse and longitudinal components of $\textbf{D}^{[2+,3+]}$ by varying the $\frac{J_{NN}^{[2+,3+]}}{J_{NN}^{[3+,3+]}}$ ratio.}
\label{S-MC4}
\end{figure*}

Which display the coupling between $\textbf{L}^{3+}$, $\textbf{M}$ and $\textbf{Q}_{K_3}$. Following these approach one can define phase $F_2$ and $F_3$. The stability of these phases strongly depends on the complex interplay between the magnetic interactions (see Fig.~\ref{S-MC2}-\ref{S-MC4}) and temperature. Hence, subtle changes in the structure is expected to drive spin-reorientation transitions.

\begin{figure*}
\begin{center}
\includegraphics[scale=0.55]{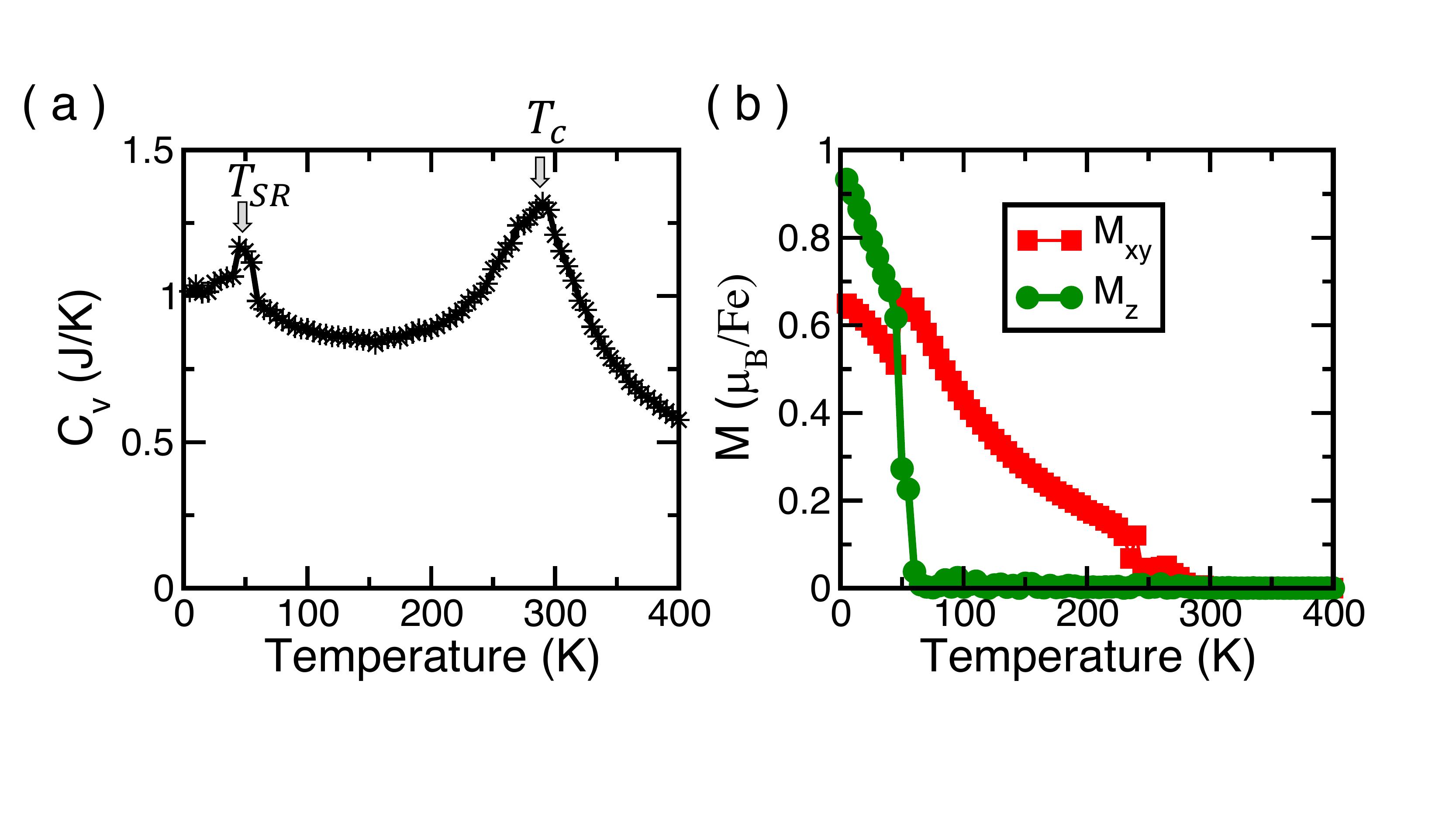} 
\end{center}
\caption{Calculated specific heat (a) and magnetization (b) as function of temperature considering $H_2$ model corresponding to the DFT estimated values of the magnetic parameters and $D^{[2+,3+]_{z}}=$ 0.3 meV and $D^{[2+,3+]_{xy}}=$ 0.7 meV.}
\label{S-MC5}
\end{figure*}

\end{document}